\RequirePackage{rotating}
\documentclass[reprint,amsmath,amssymb,aps,prstper,nofootinbib,superscriptaddress]{revtex4-2}
\usepackage{graphicx}% Include figure files
\usepackage{dcolumn}% Align table columns on decimal point
\usepackage{bm}% bold math
\usepackage{hyperref}% add hypertext capabilities
\usepackage{xcolor}
\usepackage{comment}
\usepackage{rotating}
\usepackage{colortbl}
\usepackage{enumitem}

\begin{document}

\title{\texorpdfstring{Group Dynamics in Inquiry-based Labs:\\ Gender Inequities and the Efficacy of Partner Agreements}{Group Dynamics in Inquiry-based Labs: Gender Inequities and the Efficacy of Partner Agreements}}% Force line breaks with \\

\author{Matthew Dew}
\affiliation{Laboratory of Atomic and Solid State Physics, Cornell University, Ithaca, New York 14853, USA}
\author{Emma Hunt}%
\affiliation{Department of Physics, The University of Texas at Austin, Austin, Texas 78712, USA}%
\author{Viranga Perera}
\affiliation{Department of Physics, The University of Texas at Austin, Austin, Texas 78712, USA}%
\author{Jonathan Perry}%
\affiliation{Department of Physics, The University of Texas at Austin, Austin, Texas 78712, USA}%
\author{Gregorio Ponti}
\affiliation{Department of Physics, Harvard University, Cambridge, Massachusetts 02138, USA}%
\author{Andrew Loveridge}
\email{Andrew.Loveridge@austin.utexas.edu}
\affiliation{Department of Physics, The University of Texas at Austin, Austin, Texas 78712, USA}%

\date{\today}
\begin{abstract}
Recent studies provide evidence that 
social constructivist pedagogical methods such as active learning, interactive engagement, and inquiry-based learning, while pedagogically more effective, can enable inequities in the classroom. By conducting a quantitative 
empirical examination of gender-inequitable group dynamics in two inquiry-based physics labs, we extend results of previous work. 
Using a survey on group work preferences and video recordings of lab sessions, we find similar patterns of gendered role-taking noted in prior studies. These results are not reducible to differences in students' preferences. 
We find that an intervention which employed partner agreement forms, with the goal of reducing inequities, had a positive
impact on students' engagement with equipment during a first-semester lab course. 
Our work will inform implementation of more effective interventions in the future and emphasizes challenges faced by instructors who are dedicated to both research-based pedagogical practices and efforts to promote diversity, equity, and inclusion in their classrooms.
\end{abstract}

\keywords{Diversity \& Inclusion; Epistemology, attitudes, \& beliefs; Instructional Strategies; Learning Environment; Lower Undergraduate Students}%Use showkeys class option if keyword display desired
\maketitle

\section{Introduction}
Group work is a common feature of many university physics lecture and lab courses. Research-based pedagogical practices, like active learning, interactive engagement, and inquiry-based learning, employ varying degrees of group work. Beyond its role as a pedagogical tool, group work has been identified as a learning goal itself. The American Association of Physics Teachers has designated it a component of scientific collaboration and as a learning goal for lab courses in particular \cite{AAPTlabrecs}. However, despite its potential benefits, group work introduces complexities into a course which can produce unintended effects.

Pedagogical practices which incorporate group work such as active learning, interactive engagement, and inquiry-based learning enjoy broad support within the physics education community. This support is based on a breadth of empirical studies that show them to be pedagogically more effective than traditional lecture or lab courses \cite{Freeman2014, ibrahim2022simultaneous}. However, while these practices may be best for learning overall, there is also evidence that they can enable certain inequities. For example, Quinn \textit{et al.} \cite{Quinn2020} observed that incorporating inquiry-based instructional practices into laboratory courses, compared directly with traditional labs, can result in an increase of gendered role-taking. Other studies have found this gendered division of labor to women's disadvantage \cite{Jovanovic1998,Day2016,Doucette2020}. In the context of lecture courses, Gordon \textit{et al.} \cite{Gordon2021} found that a flipped classroom had a negative impact on learning and achievement for low-income, systemically non-dominant race/ethnicity,\footnote{We use the term ``systemically non-dominant'' as explained in \cite{Jenkins2017}} and first-generation students when compared with an interactive lecture. These recent works corroborate the results of other studies, which show more generally that---absent proactive efforts by an instructor---systemically non-dominant groups engage less in active-learning components of lecture courses \cite{Aguillon2020, Dallimore2013, Lewis2019, Adams2002, Eddy2014,  Donovan2018}. Collectively, these empirical observations suggest that many popular research-based pedagogical methods can be especially susceptible to unintended inequitable outcomes. As we will review in Section \ref{sec:TheoreticalContext}, there are also sound theoretical reasons to expect this. This situation presents instructors with a serious tension: pedagogical methods which research suggests are best for overall student learning can be worse for diversity, equity, and inclusion.

Of course, this is by no means inevitable. In some contexts, research-based remediation strategies have been developed and successfully implemented \cite{Lewis2019, Dallimore2013}. It is an important goal to build upon this work, especially to include methods designed specifically for inquiry-based lab courses and lab courses more generally.

Given the integral role of group work in these pedagogical methods, it should not be surprising that one major source of these inequities lies in problematic group dynamics. The value of group work, its potential for inequities, and frameworks for promoting fair and effective group work have been important topics in physics education research for several decades, dating back to a multiyear study at the University of Minnesota \cite{Heller1992a, Heller1992b}. More recent research has explored this issue in depth in the context of laboratory courses, finding cross-cultural evidence for gendered division of labor \cite{Holmes2020}, documenting the likelihood of women to adopt secretary or project manager roles \cite{Doucette2020, Stump2023}, and assessing how the frequency of inter-group interactions is affected by lab design and group gender composition \cite{Sundstrom22}. However, these studies are limited to three institutions and it remains unclear what strategies can be used to resolve the inequities they identify.

Group dynamics in lab courses have been the subject of study beyond university physics courses. For example, important studies in university STEM courses documenting similar dynamics to the above have been conducted in engineering \cite{Camacho2011, Lewis2019}, chemistry \cite{Premo2022}, and biology classes \cite{Eddy2015, grunspan2016malesunderestimate}. In particular Donovan \textit{et al.} \cite{Donovan2018} studied different methods of group formation in a college biology class. These dynamics may contribute to higher attrition rates of women in STEM, given the role of college in the leaky pipeline \cite{pell1996fixing}. Systematic investigations of inequitable group dynamics in a pre-college setting are less common. A study by Greenfield \cite{Greenfield1997} found that girls were just as likely to manipulate equipment and record data as boys from elementary to early secondary school. Meanwhile, a study by Jovanovic  \textit{et al.} found boys manipulated equipment more than girls in grades 5-8 \cite{Jovanovic1998}. Regardless, a broad study by Burkam and Smerdon \cite{Burkam1997} emphasizes the importance of equitable equipment use for supporting women’s performance in STEM.

In this work, we present a quantitative %mixed-methods 
empirical examination of inequitable group dynamics and of a possible remediation strategy. The context of our work is two introductory physics lab courses for non-majors at a major public university. Importantly, our work is at the intersection of aforementioned results \cite{Quinn2020,Gordon2021}, given our courses' recent redesign implementing much of the inquiry-based framework advocated forin other studies \cite{Smith2021, Holmes2018} at our institution. The size and diversity of these lab courses make them useful testing grounds. The expository in \textit{The Inequality Machine: How College Divides Us} by Paul Tough \cite{Tough2021} suggests the observed student body may be of unique interest given its diversity, including dimensions such as race, socioeconomic status, and major explicitly mentioned as limitations or factors of interest in Quinn \textit{et al.} \cite{Quinn2020, UTfacts}. In this study, we do not empirically assess any link between inquiry-based course design and inequities but rather treats it as motivation for the study of the scaffolding of course elements to ameliorate them.

We choose to focus this work on the gendered aspect of group dynamics for three reasons: One, a number of previous works have also focused on gender \cite[e.g.,][]{Quinn2020, Holmes2022}, so this allows direct comparison between student populations. This is important since there is evidence that the effects of a given curriculum can depend on a student's demographics and other characteristics \cite{Eddy2017, Brownell2013}. Two, gender represents a subdivision of students with large populations in two categories. Three, if gender inequities are observed, they may signal broader inequities and provide impetus for follow up studies examining other dimensions of identity and associated inequities.

The first goal of this work is to extend previous results from Cornell University \cite{Quinn2020,Holmes2022} to the context of inquiry-based labs for non-physics STEM majors at a large public, research-intensive institution in the southern U.S., The University of Texas at Austin. Additionally, by examining student-reported preferences for group work and actual observed behaviors in a single study, we are also able to unify the previous results \cite{Quinn2020,Holmes2022} by explicitly controlling for preferences in modeling role division in lab activities. %Evidence for similar inequities despite the demographic differences will be taken to imply that they are related to shared aspects of curriculum design. Evidence for differences in preferences or in inequities will be treated as motivation for institution-specific curriculum development and intervention strategies.  
The second goal of this work is to provide an assessment of an intervention rooted in social constructivism that is meant to remediate the anticipated inequitable dynamics and which involved students completing partner agreement forms. %The intervention involved students completing an individual reflection, partner agreement forms, and partner reflection forms (see Section \ref{sec:Intervention}). Half of the lab sections examined in this study were treated as controls, while the other half were administered the intervention, allowing a side-by-side comparison.

This paper is organized as follows: In Section \ref{sec:TheoreticalContext} we provide theoretical rationale relevant to the study. In Section \ref{sec:InstructionalContext} we explain the instructional context. In Section \ref{sec:Intervention} we explain the motivation and implementation of our partner agreement intervention. The next three 
sections present the methods (Sec. \ref{sec:Methods}), results (Sec. \ref{sec:Results}), and analysis and discussion (Sec. \ref{sec:Analysis}) of the two 
major data sets collected: the preferences survey and the video observations. 
We conclude by synthesizing our results and their implications both for instruction and for future research on the dynamics of group work in physics lab courses in Section \ref{Conclusions}.

\section{Theoretical Context} \label{sec:TheoreticalContext}

\subsection{Constructivism, Instructivism, and Equity}

Many research-based instructional practices common in the physics education research literature fall under the umbrella of the learning theory known as \textit{constructivism}. Constructivism posits that knowledge is constructed by a learner through the active linking together of previous ideas or pieces of information \cite{Piaget1952}. It contrasts with ``traditional’’ teaching methods, which are based on \textit{instructivist} or \textit{behaviorist} views of learning \cite{Skinner1968} wherein knowledge is a collection of facts or skills that need to be transmitted from teacher to students \cite{Sawyer2015}. Modern pedagogical practices such as active learning \cite{Freeman2014}, inquiry-based learning \cite{McDermott1996, Smith2021}, interactive-engagement \cite{Redish1997}, and peer-assisted learning \cite{Mazur1997, GinsburgBlock2006} frequently incorporate students working in groups and are generally rooted in a more specific type of constructivism called \textit{social constructivism}. Social constructivism emphasizes that knowledge construction occurs through social interaction with people \cite{Vygotsky1978}. As reviewed in the introduction, there is empirical evidence that pedagogical practices based on social constructivism can enable inequities absent proactive remediation efforts on the part of the instructor \cite{Aguillon2020, Dallimore2013, Lewis2019, Adams2002, Eddy2014,  Donovan2018, Gordon2021, Quinn2020, Holmes2022}. Here we define, and will subsequently focus on, equity as structuring course policies that directly consider students' identities and backgrounds, so that current and past structural injustices are addressed to help students learn the course material (similar to \cite{Minow2021, DuncanAndrade2022}). Equality, on the other hand, is having course policies that treat all students the same regardless of their identities or backgrounds.

\begin{figure*}[ht]
\includegraphics[width=0.8\textwidth]{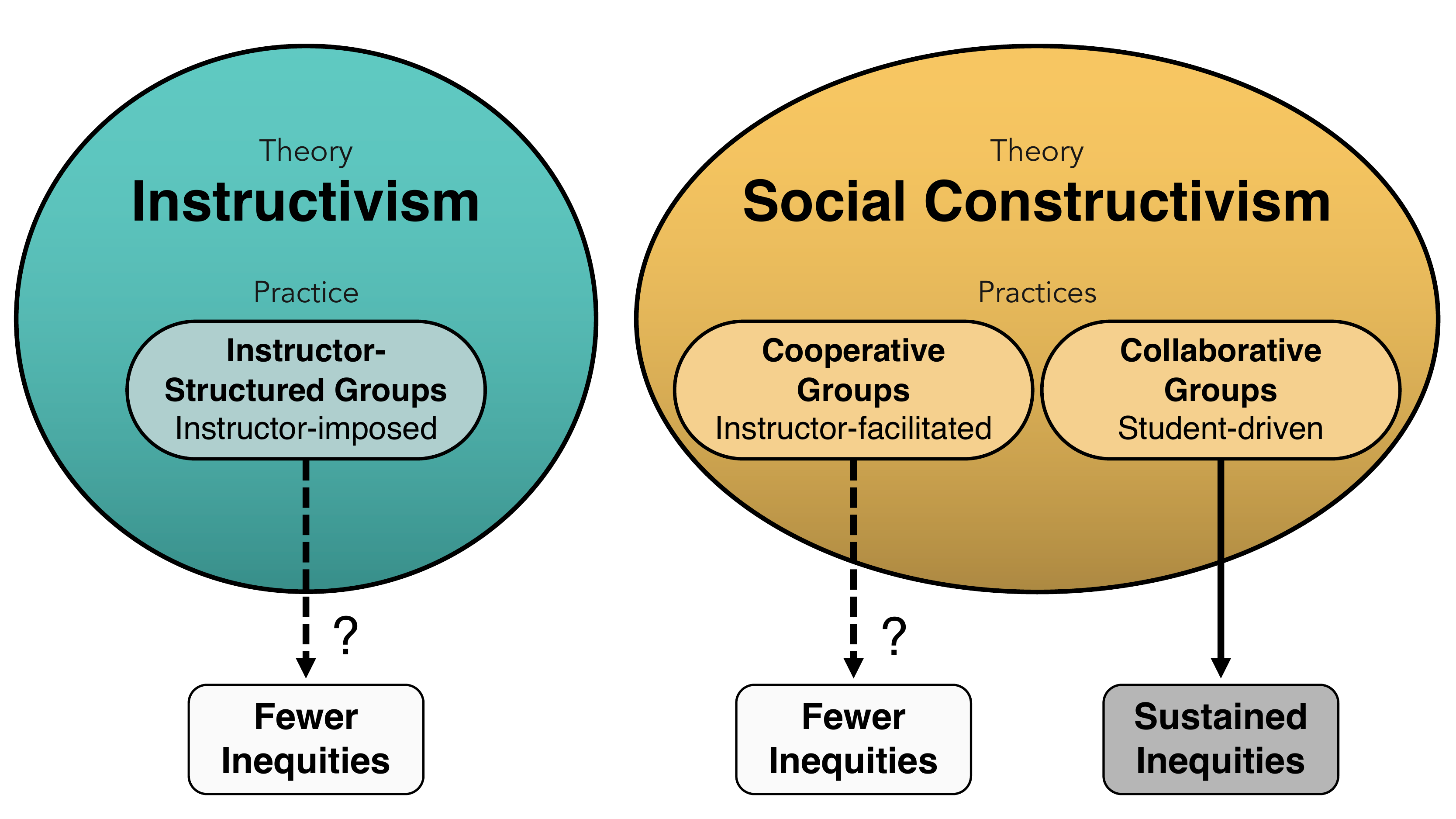}
\caption{\label{Frameworks} Pedagogical practices considered in this work depicted within their respective theoretical frameworks. For each pedagogical practice, arrows indicate known link to sustained inequities (solid line) and plausible links to less inequities (dashed lines).}
\end{figure*}

There are several theoretical reasons why we might expect some pedagogies grounded on social constructivism to have the capacity to inadvertently reinforce inequitable outcomes. A few well-studied examples which apply specifically to inquiry-based labs include:
\begin{itemize}
    \item Stereotype threat: Inquiry-based labs often require students to engage in open-ended exploration and problem-solving, which can lead to increased performance pressure. Stereotype threat, the concern of confirming negative stereotypes about one's group, can be heightened in such situations \cite{SPENCER19994}. This pressure can disproportionately affect systemically non-dominant groups, including women, and impact their performance and confidence in the lab setting.
    \item Confidence and self-efficacy: Research suggests that women, on average, may exhibit lower confidence and self-efficacy in STEM fields compared to men \cite{Ehrlinger2003, campbell1986effects, Marshman2018}. Inquiry labs can involve higher levels of autonomy, uncertainty, and risk-taking, which may lead to, or be impacted by, decreased confidence among students who are less familiar with this type of learning environment. Lower levels of confidence and self-efficacy can influence participation and engagement \cite{Linnenbrink2003}.
    \item Identity-based participation patterns: Lab classes in general involve working in groups and group problem solving. Since identity-based inequities exist in society, for example involving gender, race, class, and other categories, they can play a role governing the division of labor in group work. This division of labor can perpetuate traditional (e.g., gender) roles and create inequitable participation opportunities and experiences \cite{york2021gender, holmes2019participating}.
    \item Classroom climate and peer/instructor bias: Group work involves interaction with peers and inquiry-labs typically require proactive support from instructors. Biases, even unconscious, may influence interactions with and between students and, even inadvertently, affect the experiences and performance of systemically non-dominant groups \cite{swimming2017gender, copur2020teachersbias}.
\end{itemize}

In the case of lecture courses this issue has been studied extensively \cite{Gordon2021,Aguillon2020, Dallimore2013, Lewis2019, Adams2002, Eddy2014,  Donovan2018, Sundstrom2022b} and some specific remediation strategies have been proposed, studied, and found to be effective \cite{Lewis2019, Eddy2014}. All lab classes which involve group work, similarly, can produce inequities. Based on these theoretical considerations, we expect that inquiry-based labs would be especially susceptible to these problems, and this was found in the aforementioned study \cite{Quinn2020}, which directly compared traditional and inquiry-based labs.\footnote{It’s important to note that any extent to which traditional labs may be more equitable than inquiry labs in the absence of appropriate scaffolding of group work (or other course elements) is in an important sense superficial. Because tradional labs offer less rich and less authentically scientific activities and therefore preclude inequities via rigidity. In short, there is more room for additional tasks, social dynamics, and psychological forces in an inquiry-based lab, and so more room for inequities.} In particular, from both theoretical and empirical angles, we expect group work to be a locus of inequitable dynamics in lab courses in general and inquiry-based labs in particular.

\subsection{Group Work Practices}

To understand how students work together in groups, and how instructors might structure group work to ensure pedagogically sound and equitable outcomes, it is useful to consider three distinct group work practices: instructivist, collaborative, and cooperative learning. A visual depiction of our framework, explained below, is provided in Figure \ref{Frameworks}.

Instructivist approaches involve highly-structured group work. In theory, this maximizes instructor control over group dynamics. This may have the benefit of allowing the instructor to preclude inequitable outcomes by carefully structuring groups and how groups operate. On the other hand, by at least some measures, instructivist methods can be pedagogically inferior \cite{Freeman2014, ibrahim2022simultaneous} to other methods such as those rooted in social constructivism given they do not incorporate active involvement on the part of the student. Especially if effective group work constitutes a learning goal of a course, not just an instructional tool, we would prefer to employ a different framework for organizing group work in class.

Collaborative learning and cooperative learning, meanwhile, are rooted in social constructivism. In collaborative learning, instructors direct students to work together in groups freely, without assigning roles or structuring group work, provided they achieve a certain goal or outcome by dividing tasks effectively amongst themselves. Cooperative learning is more structured, and involves scaffolding group work so that students work together more cohesively, preferably because the goal or outcome is not achievable by individuals working independently and merely collating their work at the end. According to Davidson \cite{Davidson2021}, cooperative learning has specific characteristics: a common task or learning activity suitable for group work, small-group interaction, norms for cooperative and mutually helpful behavior among students, individual accountability and responsibility (with a possibility of including group accountability), and positive interdependence. It is important to differentiate the structure provided to students in instructivist and cooperative learning - the former dictates group structure and dynamics while the latter scaffolds students self-regulation.

Collaborative learning has the feature of being aligned with social constructivism, but it is highly plausible, and has been argued in the literature \cite{Shah2019}, that the free form nature of collaborative group work may enable social and cultural factors to preserve, reproduce, or produce inequities. Group work in inquiry-based labs without explicitly structured group dynamics resides within this framework.

Cooperative learning methods offer both alignment with social constructivism, which is better empirically supported \cite{Freeman2014, ibrahim2022simultaneous} and matches the inquiry-based learning framework of our lab courses, as well as specific structures meant to ensure healthy group dynamics. We argue it is therefore an appealing framework for resolving inequities observed in group work in inquiry-based labs. 

We will rely on instructivist, collaborative, and cooperative learning, to frame prescriptions for group work as well as our proposed intervention, as explained in the corresponding section (Sec. \ref{sec:Intervention}).

\begin{table*}[ht]
\caption{Self-reported, anonymous results from a survey of student demographic information: gender (including non-binary/other options), racial/ethnic identity \cite{NCT2015}, and parents' highest level of education \cite{NCES} %from student attitudes survey data set 
across courses ($\approx70\%$ response rate). Racial/ethnic groups were not considered mutually exclusive. Counts may not equal the total as students may not have answered all background questions or preferred to not disclose.}
\label{tab:SAGE_demo}
\begin{tabular}{p{7cm}p{2.5cm}p{2.5cm}l}
\hline
\hline
& \multicolumn{3}{ c }{Survey Responses} \\ \cline{2-4}
Student-level variables & Full sample & Physics I Lab & Physics II Lab \\ \hline
All & 1316 & 788 & 528 \\
Intervention & & & \\
\hspace{10pt} Control & 501 & 325 & 176 \\
\hspace{10pt} Partner Agreements & 427 & 243 & 184 \\
Gender & & & \\
\hspace{10pt} Women & 751 & 485 & 266\\
\hspace{10pt} Men & 479 & 255 &  224\\
\hspace{10pt} Non-binary/Other & 32 & 18 & 14\\
Race/ethnicity & & & \\
\hspace{10pt} American Indian or Alaska Native & 11 & 8 & 3 \\
\hspace{10pt} Asian & 462 & 262 & 200 \\
\hspace{10pt} Black or African American & 87 & 61 & 26 \\
\hspace{10pt} Hispanic, Latino, or Spanish & 325 & 197 & 128 \\
\hspace{10pt} Middle Eastern or North African & 42 & 26 & 16 \\
\hspace{10pt} Native Hawaiian or Other Pacific Islander & 3 & 1 & 2 \\
\hspace{10pt} White & 452 & 273 & 179 \\
\hspace{10pt} Some other race/ethnicity & 6 & 4 & 2 \\
Parents' highest level of education & & & \\
\hspace{10pt} High school & 129 & 73 & 56\\
\hspace{10pt} Some college but no degree & 82 & 61 & 21\\
\hspace{10pt} Associate's or technical degree & 50 & 33 & 17\\
\hspace{10pt} Bachelor's degree & 359 & 211 & 148\\
\hspace{10pt} Master's degree or above & 597 & 353 & 244\\
\hline \hline
\end{tabular}
\end{table*}

\section{Instructional Context} \label{sec:InstructionalContext}

We investigated two introductory physics lab courses, which took place during the Fall 2022 semester at The University of Texas at Austin. While the two courses are sequential and together constitute a two-semester introductory sequence, we studied students from the two courses simultaneously; we did not track a cohort of students through both courses. 
%The two courses are sequential, and together constitute a two-semester introductory sequence. 
Each course is a single credit hour taken by students from one of three introductory corequisite lecture sequences, including algebra-based physics, calculus-based physics for life science majors, and calculus-based physics for engineering majors. In some cases, students with prior credit are not enrolled in a corequisite lecture course. This setup mixes students from all tracks into the same lab sections %, in roughly equal proportions, 
and provides an important dimension of diversity in these lab courses.

The first course, which we will refer to here as Physics I Lab, covers standard topics in mechanics. The second, which we will refer to as Physics II Lab, continues with optics, electromagnetism, and some modern physics. Both courses are designed to implement the Structured Quantitative Inquiry framework \cite{Holmes2014a, Holmes2014b, Holmes2015}, which has some similarities to the Investigative Science Learning Environment \cite{ISLElabs} and Scientific Community \cite{Lark2014} approaches. The Structured Quantitative Inquiry framework provides students with genuine investigative freedom, supported by research-based scaffolding (i.e., invention activities \cite{Day2013}), and requires students to make fully quantitative comparisons of models with data.

Both courses are very large ($\sim$1,000 students per course) and diverse along several dimensions. At the university level, approximately 20\% of students are first-generation college students \cite{UT1st} and a similar percentage are Pell-grant eligible students \cite{Pell,UTPell}, a federal financial aid program open to students with significant financial need. This institution is also designated a Hispanic Serving Institution \cite{UTHSI}. The students in both lab courses are a representative cross-section of this student body, as shown in the demographic breakdown from an anonymous survey in Table \ref{tab:SAGE_demo}. 

Lab sections are taught by graduate teaching assistants (TAs), occasionally with assistance from undergraduate learning assistants (LAs). Each course is supervised by a faculty instructor of record and two to three graduate assistant instructors, or ``head TAs.'' Head TAs collectively are responsible for helping with curriculum development, running weekly instructional meetings, resolving grade disputes, and supporting the other TAs.

Each course includes nine lab sessions. Each session is three hours long and meets once a week. During a given lab session, students usually work in groups of two, but occasionally three. In rare instances, students work in a larger group or on their own, but that practice is discouraged. Lab activities typically involve designing an experiment to test how well a given model describes a physical system. Student groups collectively turn in a single set of informal, but structured, ``lab notes''  at the start of the following lab session which document their procedure, analysis, results, and conclusions. This gives students a week to work on analysis and writing outside of class. Because of this, most students spend class time collecting and analyzing data and leave the write-up for outside of class.

Students are allowed to pick their own partners throughout the semester. They change partners/groups every three labs, so that they work with three distinct partners or groups in a given semester. This is occasionally complicated by absences or students dropping the class, in which case groups may be slightly shuffled.

In addition to lab assignments, students individually complete pre-lab activities and a final capstone quiz or project. The pre-lab activities are completed before lab sessions as quizzes on the Canvas Learning Management System and are graded upon completion. In Physics I Lab, the final assignment is a Lab Practical Quiz that is meant to test student's mastery of essential measurement methods, analysis tools, and familiarity with equipment. In Physics II Lab, the final assignment involves students proposing and executing their own experiment on a topic of their choice and turning in a scientific poster. The inclusion of these end-of-semester individual assignments, which are worth 20\% of students' final grade, are meant to motivate individual responsibility, an important best practice for group work \cite{Johnson1994}.

\section{Intervention}\label{sec:Intervention}

Given the similar pedagogical framework and cultural context, we anticipated observing comparably inequitable group dynamics similar to previous works \cite{Quinn2020}. As such, we designed an intervention aimed at remediating expected inequities. Below we explain the motivation, form, and implementation of this intervention.

Some prescriptions for improving the equity of group work exist in the literature. For example, a highly-structured approach to designing and managing groups is advocated by Heller \textit{et al.} \cite{Heller1992a, Heller1992b}. In this framework, students are assigned specific roles which are regularly rotated, are prompted to write reflections on their experiences with their group, and are given group assignments that avoid placing women in the minority. In a summary of effective group work practices for college courses, Rosser's suggestions include ensuring rotation of instructor-defined roles throughout the semester and to avoid isolating women in groups \cite{Rosser1998}.\footnote{Rosser also recommends allowing students to select their own leader rather than having one assigned by the instructor, to allow students to assign their own roles initially, and to ensure tasks are group worthy, all of which would be categorized as cooperative group practices. Rosser's foil, the fictional professor Peter Adams, is more completely instructivist and the recommendations of the summary are largely in the direction of cooperative group practices as we advocate here.}
In the language of Section \ref{sec:TheoreticalContext}, these prescriptions may be thought of as implementations of instructivist approaches with the corresponding benefits and drawbacks.

We seek alternative solutions for a few reasons. First, we believe that teaching students to work effectively in groups is an important learning goal for lab courses in itself. This is consistent with the recommendation of the American Association of Physics Teachers, which categorizes working in small groups as a component of scientific collaboration \cite{AAPTlabrecs}. We expect that a highly-structured, top-down approach to group management rooted in an instructivist approach does not provide students with a sufficiently active role to learn to resolve problematic group dynamics. In the spirit of the ``structured inquiry'' philosophy, which aligns with social constructivism and has demonstrated effectiveness for teaching students experimental physics \cite{Walsh2022}, we seek solutions which enhance and scaffold students' active role in shaping their group work. This allows students to learn to resolve inequities and establish more effective group dynamics. Second, research has shown that highly-structured approaches to group work are often met with resistance from students \cite{Chang2018}. Third, we prefer to avoid explicit role assignment and rotation because evidence suggests it is better for student learning for them to share, not split, work, even if the splitting is equitable with respect to gender \cite{Doucette2022}. Fourth, although avoiding isolating women in groups may be sufficient to prevent inequities in collective problem solving \cite{Heller1992b}, it is unclear if this is also sufficient to prevent inequitable divisions of labor (e.g., in equipment use).

We therefore designed an intervention that was meant to give students an active role in preempting and resolving problematic group dynamics themselves. In the language of Section \ref{sec:TheoreticalContext}, we aimed to encourage cooperative group work.\footnote{Strictly speaking, if students utilize partner agreement forms to divide labor in specific ways, this may still be considered collaborative group work, but we will still judge the intervention as successful if it nevertheless reduces inequities.}
This intervention has three components:
\begin{itemize}
    \item \textbf{Individual Reflections:} Students were given a one-time, individual writing assignment, to be completed outside of class, before any lab sections met. The assignment asked students to reflect on their values and experiences with group work and to write about them. This component of the intervention is inspired heavily by the values affirmation intervention \cite{Aguilar2014, Yeager2011}, which has been shown to reduce gender achievement gaps on high-stakes exams by combating stereotype threat. The primary purpose of this exercise was to prime students' awareness of what was important for them in group work, so that they would be better equipped to recognize when their lab experience did not conform to their values for group work and learning. We expected that by providing students with an opportunity to reflect on their values, we would induce more dialogue between group members in lab and through partner agreement or reflection forms. We also speculated it would reduce stereotype threat or other identity-based issues which could play a role in group dynamics as explained in Section \ref{sec:TheoreticalContext}.
    \item \textbf{Partner Agreement Forms:} Each time students were put into a new group, they were tasked with collectively filling out a partner agreement form. This form required students to introduce themselves and to have an explicit conversation about how work would be split or shared. It was deliberately worded not to bias students towards any particular way of sharing or splitting work, while giving them an opportunity to express the preferences that were primed by the individual reflection assignment. This component of the intervention was motivated in part by results suggesting reduced inequities due to explicit conversations about equipment usage \cite{Dew2022}. It also provides a space for norm setting and the development of positive interdependence as components of Cooperative Group Work as explained in Section \ref{sec:TheoreticalContext}
    \item \textbf{Partner Reflection Forms:} Each time students returned to the same group, they were tasked with collectively filling out a reflection form. This gave them an opportunity to discuss their experience working together the previous week. This element was borrowed from the aforementioned framework from Heller \textit{et al.} \cite{Heller1992b}, since it fits with our preferred approach. Importantly, our partner reflection forms differed in that they were done collectively, not individually.
\end{itemize}
The individual reflection assignment, partner agreement form, and partner reflection form can be found in the Supplemental Materials.

Students were given participation credit for completing the individual reflection assignment. They were not given any points for completing the partner agreement nor reflection forms to minimize differences in grading across all sections. Instead, TAs required students to complete these at the start of class before proceeding with lab activities. It is worth noting that since students were not obliged to invest significant effort on these activities, some students may choose not to make maximal use of them. This was deliberate, since whether or not students proactively make use of course structures to obtain equitable and preferred modes of group dynamics is part of what we are testing in this study.

The control sections were not given the Individual Reflections, Partner Agreement forms, or Partner Reflection forms, but were otherwise treated identically to the intervention sections. It is possible that students in the control section may have learned of some aspects of the intervention going on in the sections it was applied to---indeed, we did not hide the study from the students. However, we do not expect this to have impacted our results very much. Small differences between lab sections are common due to differences in TA style and students do not perceive these differences as out of the ordinary.

Partner agreement forms have been implemented and studied before---typically in courses with a project---and these studies have found group contracts often improve communication \cite{Favela2001, Pertegal2019, Brannen2021, Chang2018}. However, few studies discuss role-taking. Students in one study were assigned a role at the start of the class \cite{Favela2001}. In another study by Chang and Brickman on group work in an introductory biology class \cite{Chang2018}, students were instructed to rotate roles explicitly as well as to write and follow group contracts. However, students did not assign or rotate roles explicitly and they often disregarded their group contracts. These implementations differ from ours as we wanted students to share their roles in a context where their work varied week-to-week. 

We assess the success of the intervention 
from the effects on equitable dynamics as observed in the video recordings.  
We expect that the intervention will eliminate or mitigate gendered role-taking and prevent introducing new inequities as compared to the control sections.

\section{Methods} \label{sec:Methods}
\subsection{Preferences Survey} \label{PreferencesMethods}
Prior to the first week of lab, students were asked to indicate their preferences for different lab activities, forms of role distribution, and leadership styles \cite{Holmes2022}. Here, the text and responses of these questions are reproduced.
\\

\noindent\fbox{
    \parbox{ \linewidth}{%
    {\footnotesize
``Which of the following experiment tasks do you prefer taking on? (Select all that apply)”
\begin{enumerate}[label=\alph*.)]
    \item Setting up the apparatus and collecting data.
    \item Writing up the lab procedures and conclusions.
    \item Analyzing data and making graphs.
    \item Managing the group progress.
    \item No preference or none of the above.
\end{enumerate}
``Which of the following approaches to group tasks do you prefer?”
\begin{enumerate}[label=\alph*.)]
    \item One where each person has a different task.
    \item One where everyone works on each task together.
    \item One where everyone takes turns with each task.
    \item No preference.
    \item Something else.
\end{enumerate}
``Which of the following approaches to leadership do you prefer?”
\begin{enumerate}[label=\alph*.)]
    \item One where one student regularly takes on the leadership role.
    \item One where no one takes on the leadership role.
    \item One where the leadership role rotates between students.
    \item No preference.
    \item Something else.
\end{enumerate}}
    }%
}
\\
\\

All three preferences questions were closed response. Multiple preferences could be selected on the activity preferences question. Only one preference could be selected on the role distribution and leadership preferences questions. These questions appeared in one of the mandatory pre-lab quizzes that students completed electronically before each lab (see Section \ref{sec:InstructionalContext}). We had 1,871 completed responses.\footnote{For the preferences survey, we use university-supplied binary data on gender since the survey in Table \ref{tab:SAGE_demo}, which allowed students to self-report their own gender, was anonymous.}

%Prior to the first week of lab, students were asked to indicate their preferences for different lab activities, forms of role distribution, and leadership styles. These questions regarding each preference appeared in one of the mandatory pre-lab quizzes that students completed electronically before each lab (see Section \ref{sec:InstructionalContext}). We had 1,871 completed responses\footnote{For the preferences survey, we use university-supplied binary data on gender since the survey in Table \ref{tab:SAGE_demo}, which allowed students to self-report their own gender, was anonymous.}. All three preferences questions were closed response. Multiple preferences could be selected on the activity preferences question. Only one preference could be selected on the role distribution and leadership preferences questions.

We examined survey results for differences across gender controlling for course, track, and the interaction of course and track in our model. The three lecture tracks act as a proxy for student majors and therefore, motivations which may in turn influence preferences. Additionally, student preferences in Physics II Lab may differ from student preferences in Physics I Lab because students have gained more familiarity with the course structure and the lab's style of group work. These changes in preferences between courses can also vary with lecture track, as some students' preferences may evolve differently to better match their priorities. These various conditions are controlled for in the logistic regression model for role preferences given in Equation \ref{eq:PreferenceRegression},

\begin{align} \label{eq:PreferenceRegression}
    \text{logit(R)} &= \beta_0 + \beta_1 \text{Woman} + \beta_2 \text{PhysIILab} + \beta_3 \text{CalcEngr} \nonumber \\
&\indent + \beta_4 \text{CalcLifeSci} + \beta_5 \text{NoCoreq} \nonumber \\
&\indent + \beta_6 (\text{PhysIILab} \ast \text{CalcEngr}) \nonumber \\
&\indent + \beta_7 (\text{PhysIILab} \ast \text{CalcLifeSci}) \nonumber \\
&\indent + \beta_8 (\text{PhysIILab} \ast \text{NoCoreq}) 
\end{align}

\noindent where $R$ is the response variable, which is the binary preference selected by students; $\beta_0$ is the intercept (Man, Physics I Lab, Algebra-based track); %$\beta_{1-8}$ %(i \neq 0)$ 
%are the model coefficients; 
Woman indicates if a student is a woman; PhysIILab indicates if a student is enrolled in Physics II Lab; CalcEngr indicates if a student is enrolled in the calculus-based physics track for engineering majors; CalcLifeSci indicates if a student is enrolled in the calculus-based physics track for life science majors; and NoCoreq indicates if a student is not enrolled in a corequisite lecture course.

For the activities preference survey, because students could select as many or as few responses as they chose, we treated each of the five responses as binary logistic regressions as in Equation \ref{eq:PreferenceRegression}. We identified gendered differences in the activities preferences using the regression estimates. For the role distribution and leadership preferences questions, because students could only select only one response, we treated each question as a multinomial logistic regression controlling for the same factors as shown in Equation \ref{eq:PreferenceRegression}. For an introduction to multinomial logistic regression and its uses, see work by Theobald \textit{et al} \cite{Theobald2019}. For these role distribution and leadership preferences questions, %we identified gendered differences using Type III ANOVA. Where statistically significant differences existed, 
we used pairwise comparisons of means to identify gendered differences in specific answer choices.

\subsection{Video Observations}
To evaluate how students divide tasks in the setting of a laboratory course, we conducted observations of recorded sections. Out of 93 lab sections covering both Physics I and Physics II Labs during the Fall 2022 semester, we video recorded four sections. These sections included one control and one intervention section for Physics I Lab, and one control and one intervention section for Physics II Lab. All sessions for the semester were recorded for these sections.

All four recorded sections were taught by ``head TA'' assistant instructors, as opposed to TAs. These sections were chosen for analysis under the assumption that they would be the most uniform subset of sections, as well as most adherent to the course objectives, minimizing instructor effects compared to using novice TAs. The four sections took place at the same time and weekday, and included two sections of the Physics I Lab and two sections of the Physics II Lab, with a control and partner agreement intervention section for each.

Labs started with a brief lecture from the TAs that ranged from roughly 10 to 30 minutes. During this period, students listened and took notes and did not start on lab work until the lecture concluded. We did not code student activities during this period. Once the TA finished their lecture, we began coding student activity. Every five minutes, a researcher coded what each student was doing at that time according to our coding scheme described below. When it was unclear what a student was doing at these five minute increments, we checked the video up to 30 seconds before and after to make a determination. %We coded the video recordings subsequent to the lecture in five minute intervals until the end of a particular lab period.

We used a coding scheme similar to Quinn \textit{et al.} \cite{Quinn2020} (see Table \ref{tab:CodingScheme} for a description of each code). The `Other' code covered a broad range of activities including students being off-task (e.g., using their phone, leaving the room, and talking to peers) as well as on-task (e.g., thinking and discussing with peers, LAs, or TAs). Students not touching but looking at a computer screen they had been scrolling on or typing at within 30 seconds or looking at a piece of paper and holding a pencil without actually writing were coded as the closest relevant activity, rather than `Other.' Additionally, when a student was holding equipment, but not using it to explicitly conduct the experiment, we coded it as `Equipment.'
%Apparently passive activities such as looking at a computer screen without scrolling or typing, holding a pencil or piece of equipment but not using it, were coded as the closest relevant activity rather than the `Other' code. % with the assumption that a student was actively engaging in that activity either before or after the relevant time-stamp. 
A student who watched another student do an activity was coded as `Other.' 

\begin{table}[ht]
\begin{ruledtabular}
\caption{Coding scheme used for video observations. The `Laptop,' `Calculator,' and `Paper' codes were later collapsed.}
\label{tab:CodingScheme}
\begin{tabular}{p{2cm}p{6.5cm}}
Code       & Description \\ \hline
Equipment  & Student was handling the equipment. This includes handling objects that are not necessarily lab equipment (e.g., a phone) when it was explicitly obvious the materials were being used to conduct the experiment (e.g., timing a pendulum's period). \\ 
Desktop    & Student was operating a lab desktop computer.  \\ 
Laptop     & Student was using a personal computer. This includes iPad or tablet use, but excludes cell phone use. \\
Calculator & Student was using a calculator. This includes cell phone use when it was explicitly obvious the phone was being used as a calculator. \\ 
Paper      & Student was using pen and paper. \\ 
Other      & All other student activities. 
\end{tabular}
\end{ruledtabular}
\end{table}
 
In our analysis, we chose to combine the `Laptop,' `Calculator,' and `Paper' codes as the distinction between these activities was unsubstantial in two important ways. First, students often used their personal computers to do calculations and take notes. Second, all three activities were associated with analysis or report-writing and required technical understanding but not physical engagement with lab materials. They were thus functionally similar to one another, but distinct from conducting the experiment itself. The `Desktop' code was not included in the grouping of `Laptop', `Calculator', and `Paper' because the desktop computer had mixed uses. Students often used the desktop computers to collect data and could therefore be linked to the `Equipment' code. Nevertheless, desktop computers were also often used to read lab instructions, conduct data analysis, or write lab notes which could align desktop computers more with the `Laptop,' `Calculator,' and `Paper' codes. Due to this conflict, we left `Desktop' as an independent code.

\subsubsection{Coders and Inter-rater Reliability}
To establish inter-rater reliability, three researchers coded 23 students in a single 3-hour recorded class session. For that purpose, we chose the second lab session in the Physics I Lab course. We chose that lab session because students frequently use a diverse set of materials and methods and it would make any difficulties with using the coding scheme apparent. For example, many students use their phones as timers in this lab session; coding this as equipment requires more careful observation than equipment observations, such as using a scale, in later labs. 
%because we consider it an especially holistic implementation of the Structured Quantitative Inquiry framework with one of the most complete representations of the intended activities found throughout both courses. 

\begin{table*}[ht]
\begin{ruledtabular}
\caption{Student and observation demographic data from the four lab sections which were recorded. Student demographic data indicates the number of men and women in each section. An observation describes one student in one lab period, thus, observation demographic data indicates the number of men and women in each session across the semester. For example, we have 8 unique students that are men in the Physics I Lab control section; we have 61 observations of these 8 unique students across the full semester due to absences.\footnote{Note that one student preferred not to share their gender.}}
\label{tab:VideoDemographics}
\begin{tabular}{lcccccccc}
      & \multicolumn{4}{c}{Student Demographics}                                  & \multicolumn{4}{c}{Observation Demographics}                              \\
      & \multicolumn{2}{c}{Physics I Lab} & \multicolumn{2}{c}{Physics II Lab} & \multicolumn{2}{c}{Physics I Lab} & \multicolumn{2}{c}{Physics II Lab} \\ \cline{2-3} \cline{4-5} \cline{6-7} \cline{8-9} 
      & Control     & Intervention    & Control     & Intervention    & Control     & Intervention    & Control     & Intervention    \\ \hline
Men   & 8           & 13              & 10          & 9               & 61          & 101             & 74          & 67              \\
Women & 15          & 12              & 8           & 10              & 127         & 103             & 65          & 69              \\
Total & 24          & 25              & 18          & 19              & 197         & 204             & 139         & 136            
\end{tabular}
\end{ruledtabular}
\end{table*}

We obtained a Fleiss' Kappa \cite{Fleiss1971} of 0.80 and Kappas over 0.75 signify excellent agreement \cite{Fleiss2003}. When we combined the `Laptop,' `Calculator,' and `Paper' codes, our Fleiss' Kappa increased to 0.84. After coding, the researchers discussed their disagreements and resolved any disputed coded segments by coming to consensus. There were no trends in which codes caused more disagreements. The researchers then coded separate sections. Two researchers (MD and AL) each coded one section of the Physics I Lab and one researcher (EH) coded both sections of the Physics II Lab. 
    
\subsubsection{Video Observations Quantitative Analysis}
While our data was coded in segments, students spent varying amounts of time in the lab. Since our research questions relate to how students work in groups, we normalized observations to a student's group. We refer to this type of data presentation as a student's ``group fraction.''

A student's group fraction for a coded activity in one class session is the fraction of codes we have of that activity out of \textit{the group's total number of codes for that activity in one class session}. In a former study, Day \textit{et al.} referred to this as ``normalized participation'' \cite{Day2016}. For a given student and class session, the group fraction for an activity ($g_{\text{activity}}$) is the number of codes of that student for that activity ($N_{\text{activity}}$) divided by the sum of the total number of codes of that activity ($N_{\text{activity}}$) over all the students in that group, as given in Equation \ref{eq:GFDef},

\begin{equation} \label{eq:GFDef}
    g_{\text{activity}} = \frac{N_{\text{activity}}}{\sum_{\text{group}} N_{\text{activity}}}
\end{equation}

For example, if in one class session we coded a student using equipment twice and their partner using it eight times, the former had an equipment group fraction of 0.2 and the latter 0.8. If we did not observe a group doing an activity for an entire class session, no student in that group was assigned a group fraction for that activity. While this is infrequent for codes such as `Equipment,' the appearance of codes such as `Desktop' varied by group and lab. 

A summary of our student population and the total number of class observations can be found in Table \ref{tab:VideoDemographics}. Most student genders were obtained in an optional supplemental survey given to students who agreed to participate in video recordings. Students were able to self-identify their gender in this survey. A total of 14 students did not fill out this survey, so we supplemented our data with university enrollment information. None of the students who filled out the survey identified as non-binary/other gender. Therefore, our data set only considers men and women.

To analyze our data, we used hierarchical linear modeling. Hierarchical linear modeling is a form of data analysis that accounts for nested structures of data. We use this form of analysis because we have repeated observations of students in each group. For an introduction to hierarchical linear modeling, see work by Van Dusen and Nissen \cite{VanDusen2019}. For our hierarchical linear model, we treat our data as a two-level model where level-1 data is a group fraction from one class session and level-2 data is a student in a particular group. This allows us to have repeated observations of a student for each group with which they work. We do not consider students across the whole semester as our level-2 data because we want to know how students act within each individual group. Additionally, we do not use a three-level model that takes groups into account as this would violate the assumption of independence of observations, as an increase in one student's group fraction necessitates a decrease in another student's group fraction.

For level-1 data, our `Equipment' group fraction for the $i$th observation of the $j$th student, $g_{\text{equipment}, ij}$, is modeled as in Equation \ref{eq:LevelOneHLMEquipment},
\begin{equation} \label{eq:LevelOneHLMEquipment}
    g_{\text{equipment}, ij} = \beta_{0j} + \beta_{1j} \text{GroupSize}  + r_{ij}
\end{equation}

\noindent where $\beta_{0j}$ is the intercept term, GroupSize is the number of group members associated with the observation, and $r_{ij}$ is the residual term. The level-2 data, which represents a student for each group they join, is modeled as in Equation \ref{eq:LevelTwoHLMEquipment},
\begin{align} \label{eq:LevelTwoHLMEquipment}
    \beta_{0j} &= \gamma_{00}  + \gamma_{01} \text{Woman} + \gamma_{02} \text{Int} \nonumber \\
    &\indent + \gamma_{03} (\text{Int}\ast\text{Woman}) +\gamma_{04} \text{EquipPref} \nonumber \\ 
    &\indent + \gamma_{05} (\text{EquipPref}\ast\text{Int}) + \gamma_{06} \text{CalcEngr} \nonumber \\
    &\indent + \gamma_{07} \text{CalcLifeSci} + \gamma_{08} \text{NoCoreq} + u_{0j} 
\end{align}

\noindent where $\gamma_{00}$ is mean intercept; Woman indicates if the student is a woman; Int indicates if the student was in a group with the partner agreements intervention; EquipPref indicates if the student preferred equipment on the preferences survey; CalcEngr indicates if a student is enrolled in the calculus-based physics track for engineering majors; CalcLifeSci indicates if a student is enrolled in the calculus-based physics track for life science majors; NoCoreq indicates if a student is not enrolled in a corequisite lecture course; and $u_{0j}$ is the residual term. For `Laptop,' `Calculator,' and `Paper' observations, the model is very similar, but the EquipPref term is replaced by two separate terms: one for a preference for notes and one for a preference for analysis (and each with a term interacting with Int). A visual and statistical check ensuring the assumptions of our hierarchical linear model are met can be found in Appendix \ref{sec:HLMAssumptions}.

Additional models accounting for group gender composition were examined as part of this study to see if this had a measurable effect on outcome accounting for preferences. We investigated this as previous research has suggested that women take on different lab roles when in mixed-gender groups \cite{Doucette2020, Quinn2020}. Statistical factors which describe the quality of this model, AIC and BIC, indicated that this additional dimension did not improve on Equation \ref{eq:LevelTwoHLMEquipment}. This may be due to lacking a sufficient number of observations from each context. %recorded sections. 
As this more complex model was not a statistical improvement, we do not include its discussion here, but note it may be valuable to examine in future studies.

\section{Results} \label{sec:Results}

In this section we briefly present key results from the preferences survey and video observations. We leave further analysis and interpretations to Section \ref{sec:Analysis}.

\subsection{Preferences Survey} \label{sec:Preferences_Results}
\begin{figure}[ht]
\includegraphics[width=\columnwidth]{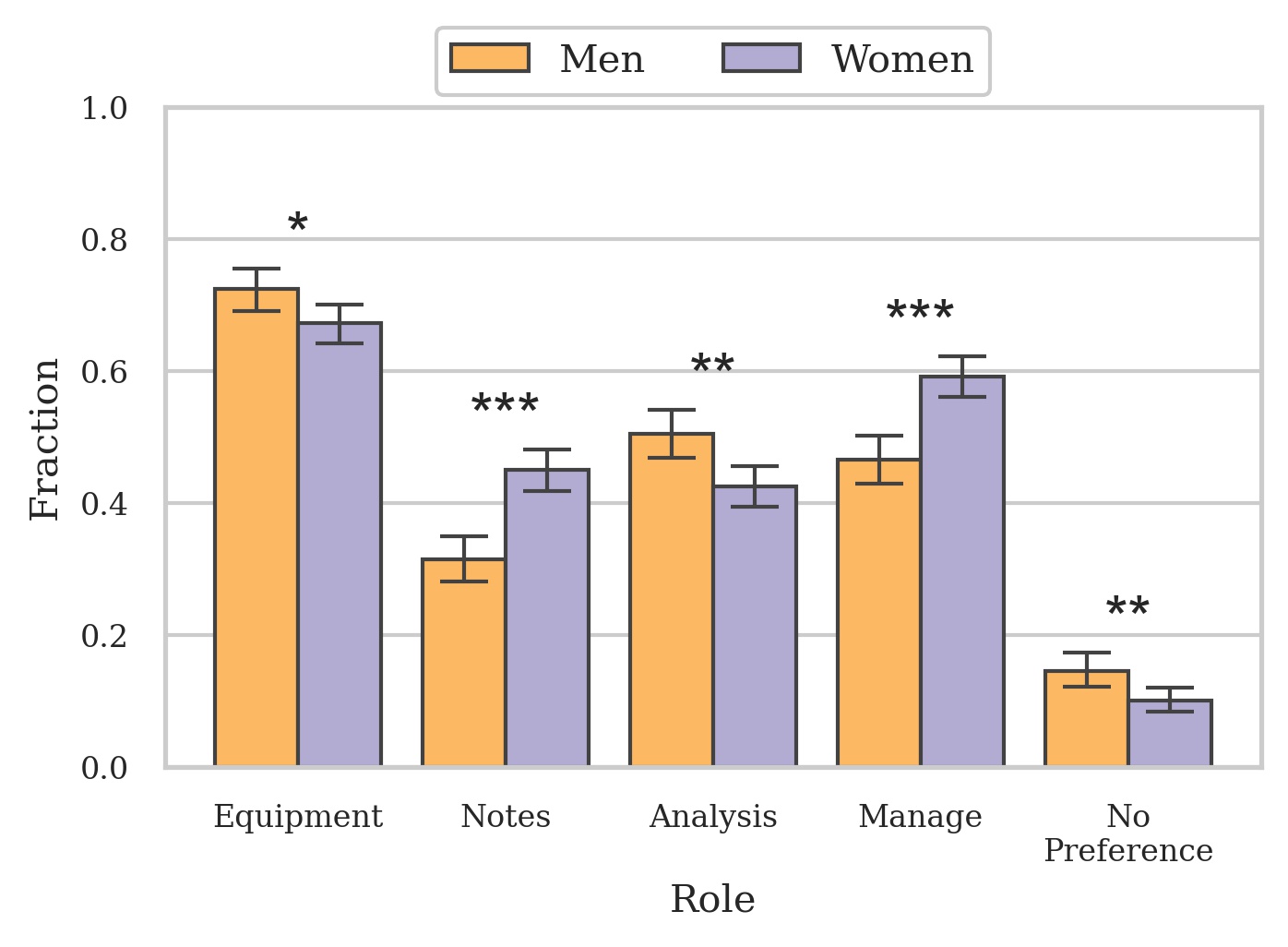}
\caption{\label{Gendered_Role_Preferences} Expected fraction of men and women that preferred a given role controlling for course, track, and the interaction of course and track. Errors bars represent 95\% confidence intervals. The asterisks denote statistical significance where $^{*}$ indicates $p<0.05$, $^{**}$ indicates $p<0.01$, and $^{***}$ indicates $p<0.001$. Students could select as many or as few roles as they wanted.}
\end{figure}

The expected fraction of student preferences for certain roles in the lab, controlling for different courses and tracks of physics that students were enrolled in, is shown in Fig. \ref{Gendered_Role_Preferences}. We find that women and men indicate different preferences at the beginning of the semester. Women more often prefer notes ($p<0.001$) and managing ($p<0.001$), while men more often prefer equipment ($p=0.024$), analysis ($p=0.001$) or have no preferred role ($p=0.004$). The full numerical results of our regression models are included in Appendix \ref{sec:survey}.

\begin{figure}[ht]
\includegraphics[width=\columnwidth]{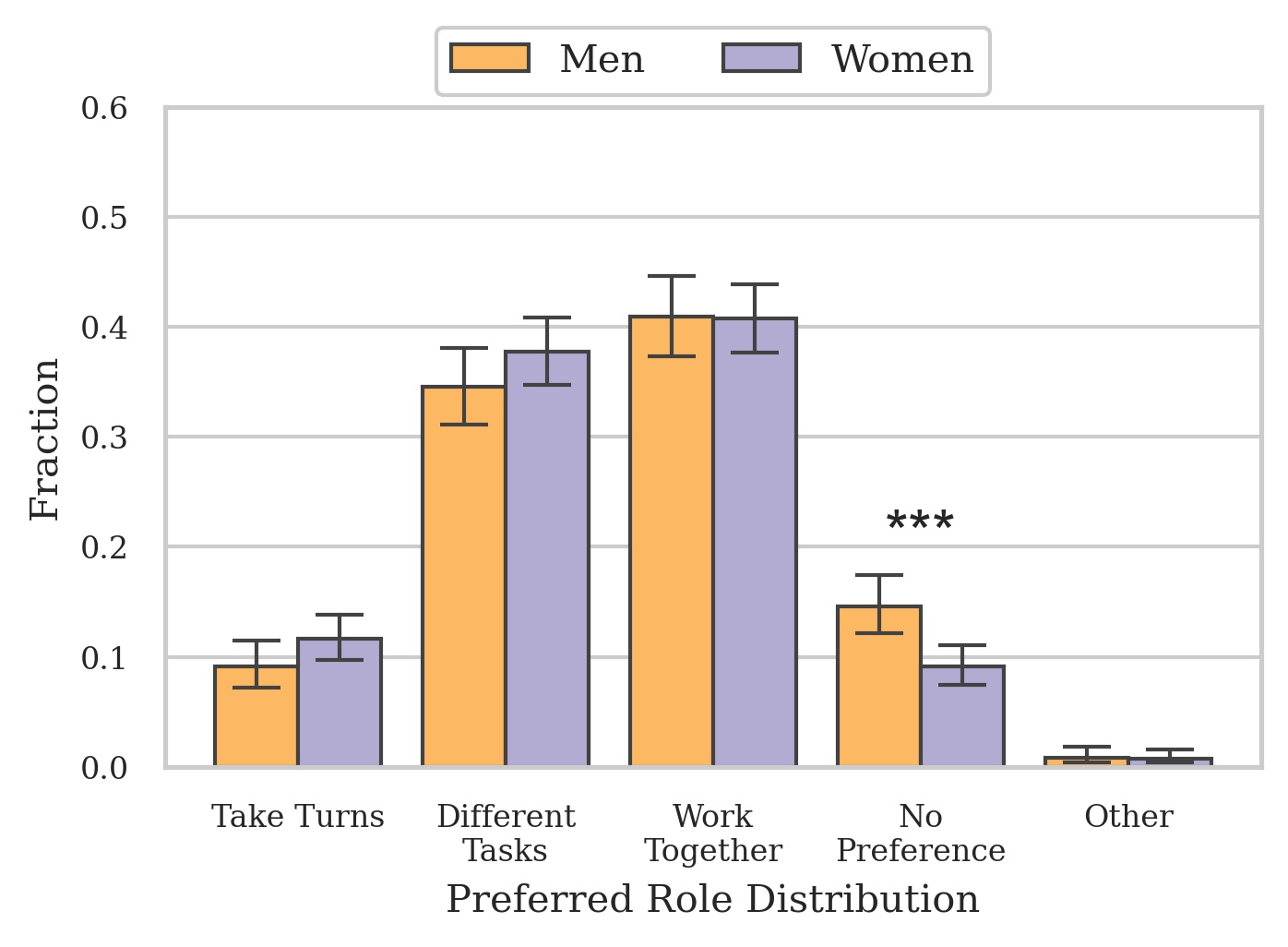}
\caption{\label{Gendered_RoleDistribution_Preferences}Expected fraction of men and women that preferred a given method of role distribution controlling for course, track, and the interaction of course and track. Errors bars represent 95\% confidence intervals. The asterisks denote statistical significance where $^{*}$ indicates $p<0.05$, $^{**}$ indicates $p<0.01$, and $^{***}$ indicates $p<0.001$. Students could select only one answer.}
\end{figure}

The expected fraction of student preferences for role distributions in lab controlling for course and track is shown in Fig. \ref{Gendered_RoleDistribution_Preferences}. %From ANOVA, we observe a statistically significant difference between men and women's preferred role distribution [$\chi^2(4) = 14.428, p=0.006$]. From a pairwise comparison of means, this is because men are more likely to report having no preference than women ($p=0.001$).
From a pairwise comparison of means, we find that men are more likely than women to report having no preference ($p=0.001$).

%Using ANOVA, we find that the interaction term between course and track is significant [$\chi^2(8) = 15.821, p=0.045$], which means we should not interpret differences in non-interacting terms (i.e. gender) further (cite).

\begin{figure}[ht]
\includegraphics[width=\columnwidth]{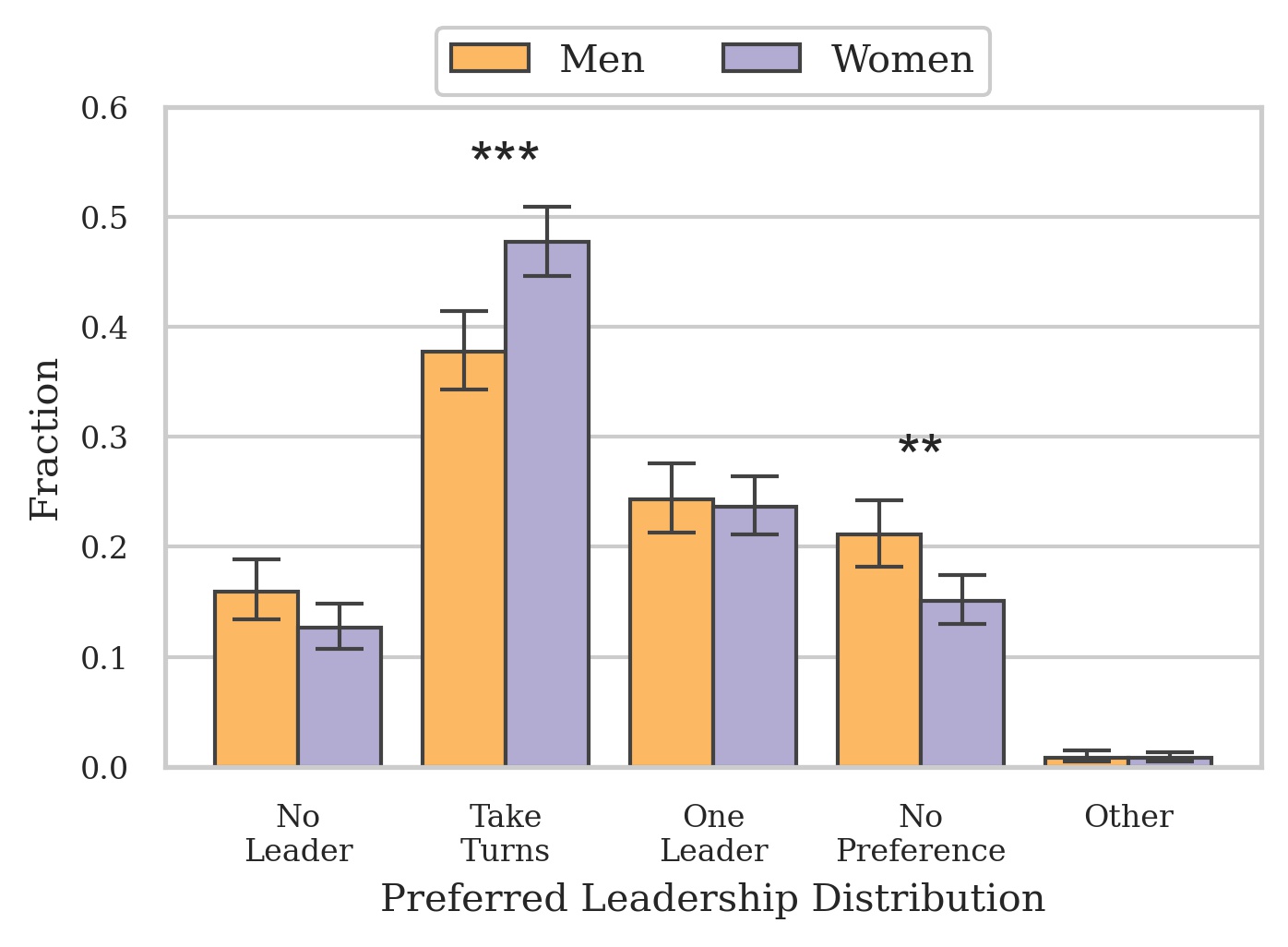}
\caption{\label{Gendered_Leadership_Preferences}Expected fraction of men and women that preferred a given leadership style controlling for course, track, and the interaction of course and track. Errors bars represent 95\% confidence intervals. The asterisks denote statistical significance where $^{*}$ indicates $p<0.05$, $^{**}$ indicates $p<0.01$, and $^{***}$ indicates $p<0.001$. Students could select only one answer.}
\end{figure}

The expected fraction of student's preferences for role distributions in lab controlling for course and track is shown in Fig. \ref{Gendered_Leadership_Preferences}. %Using ANOVA, we observe a significant difference between men and women [$\chi^2(4) = 20.438, p<0.001$]. From a pairwise comparison of means, this is because women more often reported a preference for taking turns in leadership ($p<0.001$) while men more often reported no preference ($p=0.004$).
From a pairwise comparison of means, we observe that women more often reported a preference for taking turns in leadership ($p<0.001$) while men more often reported no preference ($p=0.004$).

\begin{figure*}[ht]
\includegraphics[width=\textwidth]{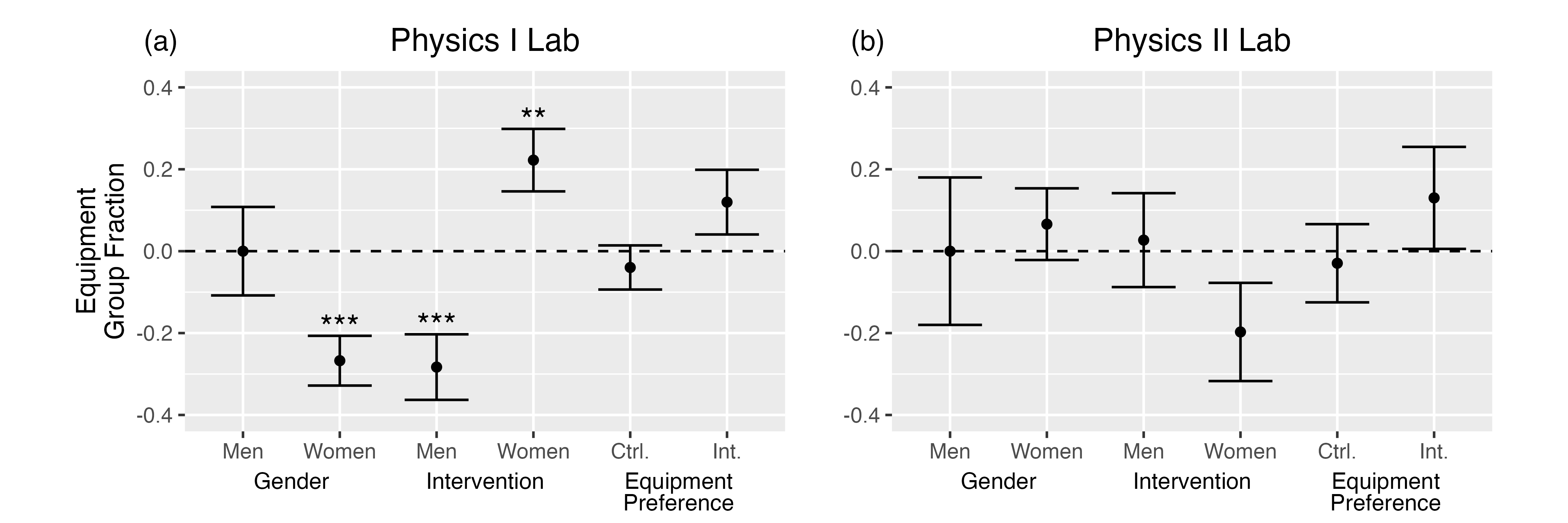}
\caption{\label{EquipmentRegression}Results from multilevel regression in (a) Physics I Lab and (b) Physics II Lab for `Equipment' group fraction. These results are controlling for group size, lecture track, and random effects. The base term for each course is the group fraction of men in the control section and enrolled in the algebra-based lecture track who did not indicate a preference for equipment. The error bars represent the standard error of the regression coefficients. The asterisks denote statistical significance where $^{*}$ indicates $p<0.05$, $^{**}$ indicates $p<0.01$, and $^{***}$ indicates $p<0.001$. The full model output can be found in Table \ref{tab:EquipmentGroupFractionRegressionSummary}.}
\end{figure*}

\subsection{Video Observations}\label{sec:Videos_Results}
%We analyze the results from `Equipment' group fraction in both Physics I and II Lab. Then, we discuss  the results from the `Laptop,' `Calculator,' and `Paper' group fraction in both course contexts.

The results of our regression model for `Equipment' group fraction are shown in Figure \ref{EquipmentRegression} and Table \ref{tab:EquipmentGroupFractionRegressionSummary}. In Physics I Lab without partner agreements, we find that women are responsible for less equipment usage than men ($\beta=-0.226 \pm 0.061, p<0.001$) accounting for group size, lecture track, equipment preference, and random effects. However, in Physics II Lab without partner agreements, we do not observe a gendered difference in equipment usage ($\beta=0.066 \pm 0.088, p=0.452$).

\begin{table}[]
\begin{ruledtabular}
\caption{Results from linear regression for `Equipment' group fraction. The table shows the regression coefficient, standard error, and $p$-value (in parentheses). %The variance explained by these models for both fixed and random effects (fixed effects only) is 42\% (16\%) for Physics I Lab and 52\% (8\%) for Physics II Lab.
The conditional (marginal) $R^2$ values for these models are 0.41 (0.16) for Physics I Lab and 0.52 (0.08) for Physics II Lab.}
\label{tab:EquipmentGroupFractionRegressionSummary}
\begin{tabular}{lcc}
Predictor                           & Physics I Lab     	& Physics II Lab  \\ \hline
Intercept                           & 1.024 $\pm$ 0.108 	& 0.790 $\pm$ 0.180  \\
                                    & ($<0.001$)	 	& ($<0.001$)         \\ \\
Group Size                          & -0.153 $\pm$ 0.031   	& -0.163 $\pm$ 0.068 \\
                                    & ($<0.001$)           	& (0.017)         \\
Track (ref: Algebra-based)                   &                   	&                 \\
\hspace{2ex}Calculus-based                   & -0.062 $\pm$ 0.059   	& 0.036 $\pm$ 0.072  \\
\hspace{8ex}for Engineers                                    & (0.288)           	& (0.615)         \\
\hspace{2ex}Calculus-based                   & 0.064 $\pm$ 0.047   	& 0.025 $\pm$ 0.094  \\
\hspace{8ex}for Life Sciences                                    & (0.172)           	& (0.792)         \\
\hspace{2ex}No Corequisite                   & 0.100 $\pm$ 0.057   	& 0.060 $\pm$ 0.089  \\
                                    & (0.083)           	& (0.496)         \\
Gender (ref: Men)                   &                   	&                 \\
\hspace{2ex}Women                   & -0.267 $\pm$ 0.061   	& 0.066 $\pm$ 0.088  \\
                                    & ($<0.001$)           	& (0.452)         \\
Partner Agreements                  &                   	&                 \\
\hspace{8ex}(ref: Control) &                 &                 \\
\hspace{2ex}Men                     & -0.283 $\pm$ 0.080   	& 0.027 $\pm$ 0.115  \\
                                    & ($<0.001$)           	& (0.814)         \\
\hspace{2ex}Women                   & 0.222 $\pm$ 0.076    	& -0.197 $\pm$ 0.120 \\
                                    & (0.004)           	& (0.101)         \\
Equipment Preferred                 &                   	&                 \\
\hspace{8ex}(ref: False)            &                       &                 \\
\hspace{2ex}Control                 & -0.040 $\pm$ 0.054   	& -0.029 $\pm$ 0.095 \\
                                    & (0.461)           	& (0.758)         \\
\hspace{2ex}Partner Agreements      & 0.120 $\pm$ 0.079    	& 0.130 $\pm$ 0.125  \\
                                    & (0.130)           	& (0.300)        
\end{tabular}
\end{ruledtabular}
\end{table}

\begin{table}[]
\begin{ruledtabular}
\caption{Results from linear regression for `Laptop,' `Calculator,' and `Paper' group fraction. The table shows the regression coefficient, standard error, and $p$-value (in parentheses). %The variance explained by these models for both fixed and random effects (fixed effects only) is 55\% (14\%) for Physics I Lab and 62\% (10\%) for Physics II Lab.
The conditional (marginal) $R^2$ values for these models are 0.55 (0.14) for Physics I Lab and 0.62 (0.10) for Physics II Lab.}
\label{tab:LCPGroupFractionRegressionSummary}
\begin{tabular}{lcc}
Predictor                         & Physics I Lab     & Physics II Lab  \\ \hline
Intercept                         & 0.821 $\pm$ 0.091 & 0.802 $\pm$ 0.157  \\
                                  & ($<0.001$)        & ($<0.001$)         \\ \\
Group Size                        & -0.138 $\pm$ 0.027 & -0.152 $\pm$ 0.062 \\
                                  & ($<0.001$)        & (0.014)         \\
Track (ref: Algebra-based)                   &                   	&                 \\
\hspace{2ex}Calculus-based                   & -0.092 $\pm$ 0.051   	& -0.075 $\pm$ 0.062  \\
\hspace{8ex}for Engineers                                    & (0.068)           	& (0.231)         \\
\hspace{2ex}Calculus-based                   & 0.032 $\pm$ 0.043   	& 0.025 $\pm$ 0.089  \\
\hspace{8ex}for Life Sciences                                    & (0.452)           	& (0.777)         \\
\hspace{2ex}No Corequisite                   & -0.043 $\pm$ 0.051   	& -0.201 $\pm$ 0.091  \\
                                    & (0.402)           	& (0.028)         \\

Gender (ref: Men)                 &                   &                 \\
\hspace{2ex}Women                 & -0.015 $\pm$ 0.056  & 0.046 $\pm$ 0.082  \\
                                  & (0.781)           & (0.578)         \\
Partner Agreements                &                   &                 \\
\hspace{8ex}(ref: Control) &                 &                 \\
\hspace{2ex}Men                   & -0.118 $\pm$ 0.070 & 0.061 $\pm$ 0.096  \\
                                  & (0.091)           & (0.524)         \\
\hspace{2ex}Women                 & 0.041 $\pm$ 0.072 & 0.083 $\pm$ 0.110  \\
                                  & (0.565)           & (0.449)         \\
Notes Preferred      		  &                   & 		\\
\hspace{8ex}(ref: False)          &                   &                 \\
\hspace{2ex}Control               & -0.014 $\pm$ 0.046 & -0.002 $\pm$ 0.077  \\
                                  & (0.761)           & (0.978)         \\
\hspace{2ex}Partner Agreements    & 0.119 $\pm$ 0.069 & -0.144 $\pm$ 0.113 \\
                                  & (0.083)           & (0.205)         \\
Analysis Preferred 		  &		      &                 \\
\hspace{8ex}(ref: False)          &                   &                 \\
\hspace{2ex}Control               & 0.023 $\pm$ 0.047 & 0.043 $\pm$ 0.077  \\
                                  & (0.632)           & (0.577)         \\
\hspace{2ex}Partner Agreements    & -0.004 $\pm$ 0.065 & -0.068 $\pm$ 0.103 \\
                                  & (0.949)           & (0.509)        
%\multicolumn{3}{r}{* $p<0.05$, ** $p<0.01$, *** $p<0.001$}                                 
\end{tabular}
\end{ruledtabular}
\end{table}

The partner agreements had differing effectiveness across the two courses. In Physics I Lab, compared to men in the control section, men in the partner agreements section had a lower average `Equipment' group fraction ($\beta=-0.283 \pm 0.080, p<0.001$), while women had a higher average `Equipment' group fraction ($\beta=0.222 \pm 0.076, p=0.004$). In Physics II Lab, compared to men in the control section, we did not observe a statistically significant difference in `Equipment' group fraction for men or women who used partner agreements. %we find women who used partner agreements had a lower average `Equipment' group fraction ($\beta=-0.206 \pm 0.102, p=0.046$) while partner agreements did not change men's average `Equipment' group fraction. 

Notably, indicating a preference for using equipment in Physics I Lab did not lead to a statistically significant increase in `Equipment' group fraction for students with or without partner agreements. In Physics II Lab, similarly, indicating a preference for equipment did not lead to a statistically significant increase in `Equipment' group fraction for students with or without partner agreements.

%Notably, indicating a preference for using equipment in Physics I Lab did not lead to an increase in `Equipment' group fraction for students without partner agreements, however, it did for students with partner agreements ($\beta=0.150 \pm 0.074, p=0.045$). In Physics II Lab, however, indicating a preference for equipment did not lead to a statistically significant increase in `Equipment' group fraction for students with or without partner agreements. 

\begin{figure*}[t]
\includegraphics[width=\textwidth]{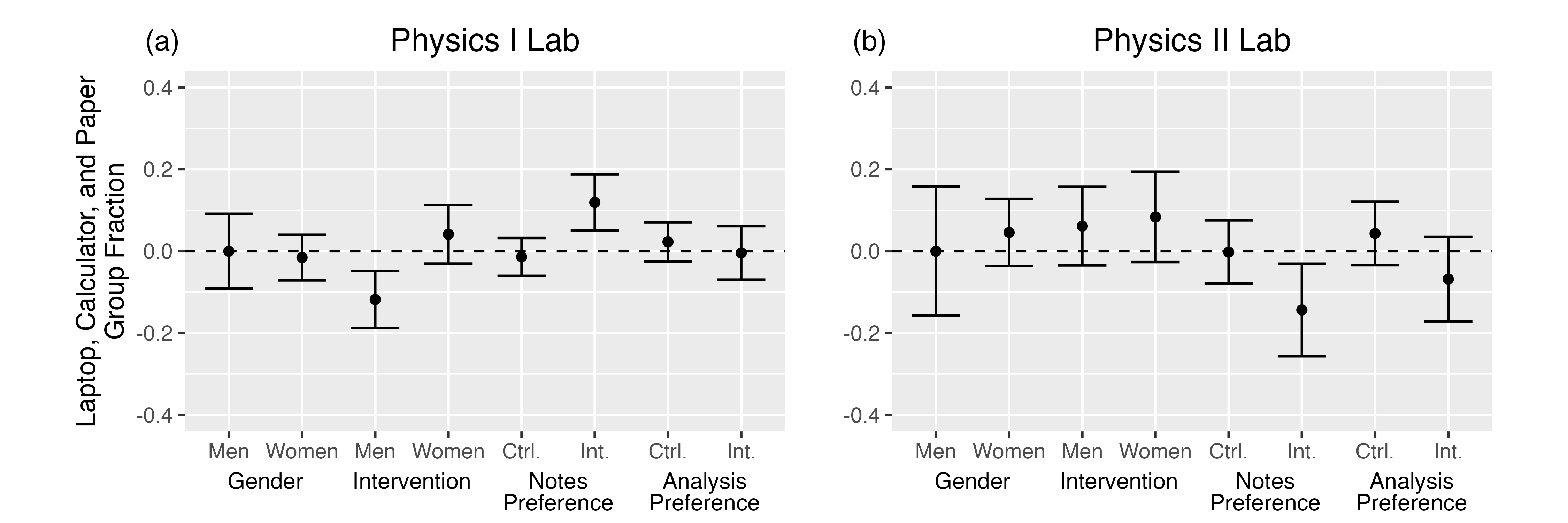}
\caption{\label{LCPRegression}Results from multilevel regression in (a) Physics I Lab and (b) Physics II Lab for `Laptop', `Calculator', and `Paper' group fraction. These results are controlling for group size, lecture track, and random effects. The base term for each course is the group fraction of men in the control section and enrolled in the algebra-based lecture track who did not indicate a preference for notes or analysis. The error bars represent the standard error of the regression coefficients. There were no statistically significant differences. The full model output can be found in Table \ref{tab:LCPGroupFractionRegressionSummary}.}
\end{figure*}

Figure \ref{LCPRegression} and Table \ref{tab:LCPGroupFractionRegressionSummary} show the results of our regression model for `Laptop,' `Calculator,' and `Paper' group fraction. In both Physics I Lab and Physics II Lab without partner agreements, we do not see any statistically significant difference in `Laptop,' `Calculator,' and `Paper' group fraction between men and women. We also see no statistically significant effects from partner agreements on men or women's `Laptop,' `Calculator,' and `Paper' group fraction. Similarly, we see no statistically significant effects from student preferences on `Laptop,' `Calculator,' and `Paper' group fraction in either course with or without partner agreements. 

\section{Analysis \& Discussion} \label{sec:Analysis}

In this section we build on the results presented briefly in the previous section. We analyze and interpret these results and models in the context of our theoretical lens and research questions, presented earlier in the paper.  

\subsection{Preferences Survey} 
When surveyed at the beginning of a semester, the most popular lab activity among students was equipment use. Men indicated a preference for equipment usage slightly more often than women. The difference in the expected fractions between men and women resembles the magnitude of the difference found by Holmes \textit{et al.} \cite{Holmes2022}, although they did not conduct tests of statistical significance on their data set. Men were also more likely to prefer the analysis role or to indicate having no role preference. Women more often expressed a preference for note-taking and managing at the start of the semester. This may be related to the ``Hermione’’ and ``secretary’’ archetypes from Doucette \textit{et al.} \cite{Doucette2020}, suggesting that previously observed gendered division of labor may be driven in part by student preferences. %Although the coded video data does not capture management activities, it does include note-taking (through `Laptop,' `Calculator,' and `Notes'), so this may provide context for the next section as well.

For student preferences in role distributions, men and women in our courses have similar preferences, albeit men are more likely to have no preference. 
%For students preferences in role distributions we cannot comment on statistically significant differences due to gender because our interaction term had a statistically significant impact. We will only comment on the general trends.
Generally, students nearly equally prefer working on different tasks or working together on the same tasks. %To borrow language from Doucette et al. \cite{Doucette2022}, students nearly equally prefer ``splitting'' and ``sharing'' the work. 
In the language of Section \ref{sec:TheoreticalContext}, this could suggest that students are similarly likely to prefer collaborative or cooperative modes of group work. This is notably different from the findings of Holmes \textit{et al.} \cite{Holmes2022}, where both men and women preferred working on the same task together in a laboratory class targeted towards physics majors. This comparatively strong preference among students in our study for splitting the work may be because these students, not being physics majors as was the case in Holmes \textit{et al.} \cite{Holmes2022}, prioritize efficiently completing lab work over content mastery. Another difference with previous results appears in the leadership preferences. In Holmes \textit{et al.} \cite{Holmes2022}, students were unreceptive to a singular leader and were comparatively more likely to prefer having no leader. A large fraction of our study's students want some form of leadership, whether that is a rotating or singular leader. %This difference could again be a product of students' expressed desire for efficiency.

The observations of the last paragraph have important
implications, since student preferences
inevitably intermix with course structures and interventions
to produce outcomes. Differences in student
preferences between populations suggest best practices
for group work may require some institution-specific
or course-specific tailoring.

%The observations of the last paragraph have two important implications. First, since student preferences inevitably intermix with course structures and interventions to produce outcomes, differences in student preferences between populations suggest best practices for group work may require some institution-specific or course-specific tailoring. Second, if large subsets of students prefer collaborative---rather than cooperative---group work, they may use partner agreements to cement a collaborative group structure. %if large subsets of students prefer modes of group work, such as splitting the work, which are not optimal for student learning, 
%then interventions meant to promote effective group work which rely on student agency may be of limited efficacy. 

\subsection{Video Observations}\label{sec:AnalysisVideo}
\subsubsection{Control and Partner Agreements}
In the non-intervention sections, we observed men being responsible for more equipment usage in their groups than women in Physics I Lab, but not in Physics II Lab. However, we did not find gendered differences in how men and women used laptops, calculators, and paper in their groups. Students primarily used their laptops for data analysis and note-taking, while they near-exclusively used calculators for analysis and paper for notes. This suggests that, in terms of roles, men were more likely to be a group's equipment user. We cannot claim that men or women are more or less likely to be note-takers or data analysts.

%Across all sections and throughout the semester, we found that men spent more of their class time using equipment and were responsible for more of their group's equipment usage than women. Similarly, we found that women spent more of their class time on laptop, calculator, and paper and were responsible for these activities more often than their group mates. Students primarily used their laptops for data analysis and note-taking, while they near-exclusively use calculators for analysis and paper for notes. This suggests that, in terms of roles, men were more likely to be a group's equipment user and women were more likely to be a data analyst and/or note-taker. 

Men being more likely to be their group's equipment user echoes the results of previous studies that have examined student roles in physics labs. %These results echo results of previous studies that have examined student roles in physics labs. 
In observations of similarly structured inquiry-based physics labs, Quinn \textit{et al.} \cite{Quinn2020} found that men were more often responsible for equipment usage, however, they also found women used laptops more. Another study of the labs at the same university found that equipment usage was similarly gendered for in-person courses, but that online courses with fixed groups across the semester resolved this inequity \cite{Dew2022}.

We found differences compared with a study by Day \textit{et al.} \cite{Day2016} who analyzed role distribution among mixed-gender pairs of students. While their coding system differed from ours in that they had codes just for equipment, computer, and everything else (also called `Other'), they found that men and women used equipment a similar amount. However, they observed men more frequently using computers and women were more frequently coded `Other.' Since students submitted notes on paper in that course, this suggests that men in the course did more data analysis while women did more note-taking. 

Recall that in Section \ref{sec:Intervention} we provided a consideration for assessing the effectiveness of the intervention. %First and foremost, 
It is effective if it resolves or at least mitigates any gender inequitable division of roles and does not introduce any new gendered inequities. %Second, it is effective if it encourages sharing of roles, rather than splitting. The latter is evident when distributions become more compact, as that means group members are doing tasks a similar amount.
By this metric, the results are positive %mixed 
regarding the effectiveness of the partner agreement intervention.

Our results suggest that the partner agreements led to more equitable equipment usage among students in Physics I Lab. For `Laptop,' `Calculator,' and `Paper' in Physics I Lab, the partner agreements did not alter the gendered distribution of labor.
%When we consider Physics II Lab, however, the intervention introduces gender inequitable equipment usage. For `Laptop,' `Calculator,' and `Paper' in Physics II Lab, there were no statistically significant shifts.
Similarly, when we consider Physics II Lab, we see no statistically shifts for `Equipment' group fraction or `Laptop,' `Calculator,' and `Paper' group fraction.

While partner agreements do not explicitly prompt students to consider gender equitable labor, their effectiveness for `Equipment' usage in Physics I Lab has a possible explanation.
%Given this mixed effectiveness of partner agreements, it is of note that `Equipment' usage is more equitable in Physics I Lab. There are a few possible explanations for this. 
In Physics I Lab, students are required to complete a practical quiz at the end of the course which tests, among other things, skills with equipment. 
%Physics II Lab, however, does not have a practical quiz; it has a final project. While students may gain familiarity with note-taking and data analysis in various other courses, the equipment they use in Physics I Lab is likely not present in their other courses. Therefore, getting experience with equipment to prepare for the practical quiz might make that role particularly salient for students. 
When completing partner agreements, students may be more motivated to ensure everyone gets equal experience with the equipment or more motivated to self-advocate for a role in equipment use. 

Zhang \textit{et al.}'s findings on the effects of formal contracts and competence trust on group work might offer an additional lens on the outcomes observed here \cite{Zhang2018}. Competence trust is how confident a student is in their partners’ experience and ability to complete a task at a high level \cite{Jeff2000}. Zhang \textit{et al.}'s findings show maximal benefit to group work when group contracts are present and groups have mild competence trust, where neither partner is perceived to be notably more or less capable by the other. In the courses studied here, competence trust may be activity-specific; students may inherently perceive some others as more or less competent in Physics I Lab due to their experiences with and perceptions of the subject, equipment, or processes. In Physics II Lab, however, students may perceive each other as being more uniformly capable for two reasons. They may perceive everyone as \textit{less} capable due to the sophisticated, and unfamiliar, electronics in Physics II Lab. Alternatively, they may perceive everyone as \textit{more} capable because they all completed Physics I Lab and have gained familiarity with group work in a university physics lab context. Similarly, students in Physics II Lab may have more college experience or higher maturity levels which could impact students' perceptions of each other's capabilities and influence the expression of their own interests.

%differ between the Physics I and Physics II Labs. This is plausible since the equipment used in Physics II Lab often involves sophisticated electronics compared with the higher frequency of everyday or otherwise familiar items in Physics I Lab.

Despite students not being given a grade incentive, the changes in equipment usage from the intervention suggest students are engaging with and using the partner agreements. It is possible that students use the conversations prompted by the partner agreements as a tool to complete group work more efficiently, but the results noted here show a more equitable outcome -- at least for equipment usage -- as a result of their inclusion in the course.
%Students may use the conversations prompted by the partner agreements to make more effective use of their time; they might view it as an efficient way to prepare for lab work.

\subsubsection{Role Preferences}
Across the Physics I and Physics II Labs, we found that student preferences at the start of the semester did not correlate with their frequency of engaging in observed lab activities. So students who identified as preferring to use equipment more were not later observed to use equipment more often. Interestingly, this behavior was consistent across both control and partner agreement sections.
This common trend of student preferences not affecting actual roles suggests students are dividing roles for other reasons. Previous research has found that students often informally take on lab roles \cite{Quinn2020, Holmes2022}. This suggests that informal role assignments are not even due to students with proclivities towards certain work instinctively taking it up. In their study of a project-based physics lab, Stump \textit{et al.} \cite{Stump2023} found a managerial role allowed some women a form of self-expression which could promote identity development, engagement, and learning. If students are not taking on roles they want to do, it could potentially impact their learning outcomes and affect (e.g. attitudes).

\subsection{Limitations}\label{Limitations}
While student preferences for roles were probed in the survey, our observation protocol did not directly observe students within all of these roles. Notably, we did not directly observe note-taking and data analysis; we observed students using laptops and calculators and writing on paper. Because these activities encompass both note-taking and data analysis, we can not fully glean how students divided roles. It is possible that some students tended to take on more secretarial or lead scientist roles, however, we could not observe these differences. 

It is also important to note that the neither the manipulation of equipment nor note taking should be equated with the entirety of scientific practice. Both are necessary components, and they are aspects which are learning goals of the course which have corresponding graded assignments (like the lab notes for each lab or the end of the semester lab practical and final projects). But other activities beyond the scope of our video coding scheme such as problem-posing, discussions of experimental design, and data analysis, are also important components. We do not, therefore, have a complete portrait of the inequities in this lab course, even though the differences in equipment usage which are not reducible to preferences implies inequities for at least this aspect according to the definition adopted in Section \ref{sec:TheoreticalContext}.

Since the interventions were applied to separate sections and implemented for an entire semester, the data compares sections with potential for different random effects. Hierarchical linear modeling accounts for random differences in behavior of students across their sessions with one group, however, there are possible random differences in sections. We could not account for these because we only have one section per condition (i.e., course and control/intervention). We attempted to minimize the differences between sections by analyzing sections taught by head TAs occurring on the same weekdays at the same times; %at the same time of day; 
uncontrollable factors can cause otherwise identical sections to differ in significant ways which could have altered the apparent intervention effectiveness. %This means differences in initial conditions other than the intervention itself could have affected the apparent intervention effectiveness. %For example, student personalities, gender composition, etc. all differed between each section. 
TA behavior and identity also differed between sections. This could have had an impact on the intervention's effectiveness. 
%an effect that is as large as an intervention. 
However, previous research finds little impact due to instructor gender, suggesting this is unlikely \cite{Hoffmann2009, Solanki2018, Dew2021, Ozmetin2021}, although this remains an open question.

In this analysis we have neglected to discuss a ``group manager'' role. This is because observations were conducted on video that lacked audio. This put identifying a group manager outside the possible scope of this study. Other studies have analyzed the positive \cite{Stump2023} and negative \cite{Doucette2020} aspects group management can have on a woman's experience in physics labs. 

We have not controlled for the possibility that students may have chosen to work with students with whom they were previously acquainted with and acknowledge that this could play a role in outcomes. Group formation procedures were the same between control and intervention sections, so this does not impact the assessment of our intervention in its context, but this is an important factor worth considering in other contexts or future studies. See Pulgar \textit{et al.} \cite{PhysRevPhysEducRes.18.010146} for one example of a study investigating this dynamic.

We have only considered groups of sizes two and three. Other lab courses may have larger groups. Since this affects the available number of roles per student, it could conceivably produce different outcomes.

\section{Conclusions}\label{Conclusions}

In this section, we summarize our findings holistically in light of our research goals, draw some conclusions, indicate implications for instruction, and suggest avenues for future work.

The first goal of this study was to apply the same methods as previous work \cite{Quinn2020, Holmes2022} in the context of our courses. Given the similarity in course framework, we expected we would observe similar inequities. We did, indeed, observe inequities \cite{Quinn2020} and they do not appear to be reducible to differences in lecture tracks or role preferences among our students \cite{Holmes2022}. Given the differences in student populations between those studies and ours, we take this as evidence that they are common to inquiry-based labs or lab courses more broadly. Importantly, by including both preferences and observations in a single study, and by including preferences as a component of our model, we unify and therefore strengthen the conclusion of these studies that preferences do not account for differences in observed lab activities. In fact, we found role preferences played very little part at all in determining actual role-taking in labs, which further suggests students may require scaffolded group work to ensure their participation reflects their interests and/or is more equitable.

This emphasizes the tension between best practice pedagogical methods and efforts to promote diversity, equity, and inclusion. An important point of Quinn \textit{et al.} \cite{Quinn2020} is that these inequities are not merely background effects of culture on all physics courses, they are consequences which can be inadvertently reinforced by particular choices of curriculum design.

Although we did see some relatively small pre-semester gendered differences in preferences that match Holmes \textit{et al.} \cite{Holmes2022}, we did find some differences in preferences between the two student populations. Our students are more likely to prefer dividing tasks and a single leader. Our observations, however, indicate that student preferences do not result in statistically significant differences in observed behavior. In the language of Section \ref{sec:TheoreticalContext}, this suggests that students are willing to engage in cooperative group structures with a brief, recurring intervention that does not explicitly compel them to equitably divide group tasks.

%Since student preferences could play a role in driving outcomes, this means solutions to inequities which prove effective in one context may be less effective in another. For example, if our interpretation is correct that our students (comparatively) value efficiency, this should be considered when designing an intervention. Students may require additional motivation to master the material before they are willing to expend effort to resolve inequitable group dynamics. In the language of Section \ref{sec:TheoreticalContext}, students may prefer collaborative rather than cooperative group work and this may limit their participation in or utilization of scaffolding meant to promote cooperative group work.

In this study we have only examined inequities based on student gender. There may be inequities based on other demographic criteria such as race/ethnicity or students' academic backgrounds \cite{Gordon2021}. We plan to follow up with future work %using our data set 
to explore these possibilities. Given the differences between Physics I Lab and Physics II Lab, it would also be interesting to conduct a future study in which the preferences survey is administered both pre-semester and post-semester to observe how lab activities may influence preferences. Additionally, a future study could examine preferences and observations within each group to explore if preferences have different effects at the start of the semester or the first sessions of new lab groups.

The second goal of this study was to implement and evaluate the impact of an intervention meant to reduce inequitable task division. This intervention was intended to work by scaffolding group work to enhance student's active role in shaping group dynamics, ideally producing something closer to cooperative rather than collaborative group work. %From multiple points of view
%Our results show that the intervention had mixed results or was ineffective, since observations show mixed or even counterproductive trends in equitable task division.
Our results show that the intervention had positive results when students had motivation (e.g. a summative practical quiz) to self-advocate for an equal role in equipment usage, since observations showed improvements where inequitable task division existed.

%It may be that students simply do not perceive these inequities. If students are not aware of or are unconcerned with inequities, we would not expect our intervention to be successful, since it relies on students to use the intervention as an opportunity to address them. Results from surveys or interviews of students may shed light on this and we plan future work examining this issue.

While our results are promising that small interventions can help mitigate inequitable group dynamics, further study is needed to investigate factors noted above, and to examine if these outcomes are true in other contexts. In particular, developing interventions to further instantiate students' self-interest in exploring all roles in a lab course would be beneficial. Using these interventions to broaden student awareness of best practices may improve their intrinsic motivation and promote equitable and effective learning experiences.

%Our results call for new or better intervention strategies. One possibility is to alter the partner agreement forms to include explicit prescriptions for research-based best practices of groups for the sake of learning (e.g. sharing rather than splitting) and to meanwhile better communicate and further instantiate the students’ self-interest in mastering all roles in the lab course. Students who are more aware of best practices and more motivated to implement them may make better use of agreement and reflection forms.

A motivation for this work were the theoretical reasons and empirical evidence that inquiry-based labs may enable or inadvertently reinforce inequities. But we have not tested this relationship in this work and, importantly, our conclusions may apply more broadly to traditional labs as well. In fact, our results emphasize the importance of the distinction between inquiry-based labs as a general laboratory design strategy and the structure and scaffolding of group dynamics (among other aspects of the course). Although there may be reasons to believe that inquiry-based design have the potential to enable inequities, our results show that this can be affected by how group dynamics are managed by instructors. Because inequities exist in society, instructors need to intentionally design lab experiences that scaffold and direct groups so as to create equitable experiences regardless of other curriculum choices. Indeed, the richer and more authentically scientific lab activities are, the more this is likely to be both necessary and valuable.

\section*{Acknowledgements}
We would like to thank Natasha Holmes, Kathryn Hendren, Jane Huk, Vernita Gordon, and John Yeazell for many useful discussions and feedback. We would like to thank Ben Costello (and his supply room workers) for implementing the video recording hardware and managing the software and servers. %We would like to thank Laura Costello for feedback on our post-semester survey. 
We would also like to thank the College of Natural Sciences at The University of Texas at Austin for providing funding for this project through the 21st Century Curriculum Redesign Effort, and Kristin Patterson and Keely Finkelstein for their assistance with applying for and administering these funds. We would also like to thank the many students who agreed to participate in this study and the graduate student teaching assistants and undergraduate learning assistants who assisted with teaching the courses.
%\newpage
\appendix

\section{Authorship Matrix}
Due to the complexity of this project, frequently used author ordering conventions seemed inadequate to properly recognize contributions from all authors. The author ordering of this paper is alphabetical, with an anchor author for the principal investigator. The matrix shown in Table \ref{tab:AuthorMatrix} describes each author's primary or secondary contribution to significant parts of this project from inception to publication.

\begin{table}[h]
\begin{ruledtabular}
\caption{Contributions of authors. Dark shading corresponds to a major contribution, while a lighter shading corresponds to a minor contribution.}
\label{tab:AuthorMatrix}
\begin{tabular}{lcccccc}

 & MD & EH & AL & VP & JP & GP \\ \hline
Concept \& Project Proposal & & & \cellcolor{black!75} &  & \cellcolor{black!25} & \cellcolor{black!75} \\
%Pilot Study (*) & & & & & & \\
Study Design \& IRB & & & \cellcolor{black!75} & \cellcolor{black!25} & \cellcolor{black!25} & \cellcolor{black!75} \\
Funding & & & \cellcolor{black!75} &  & \cellcolor{black!25} & \\
Data Collection & \cellcolor{black!25} & \cellcolor{black!25} & \cellcolor{black!75} & & & \cellcolor{black!75} \\
Video Coding & \cellcolor{black!75} & \cellcolor{black!75} & \cellcolor{black!25} & & & \\
Analysis & \cellcolor{black!75} & \cellcolor{black!25} & \cellcolor{black!25} &  & \cellcolor{black!25} & \\
Theory \& Literature Review & \cellcolor{black!25} & \cellcolor{black!25} & \cellcolor{black!75} & \cellcolor{black!75} & \cellcolor{black!25} & \\
Project Management & & & \cellcolor{black!75} &  & \cellcolor{black!75} & \\
Writing \& Editing & \cellcolor{black!75} & \cellcolor{black!75} & \cellcolor{black!75}& \cellcolor{black!75} & \cellcolor{black!75} & \cellcolor{black!75} \\

\end{tabular}
\end{ruledtabular}
\end{table}

\section{Assumptions for Hierarchical Linear Modeling} \label{sec:HLMAssumptions}
\begin{figure}[h]
\includegraphics[width=\columnwidth]{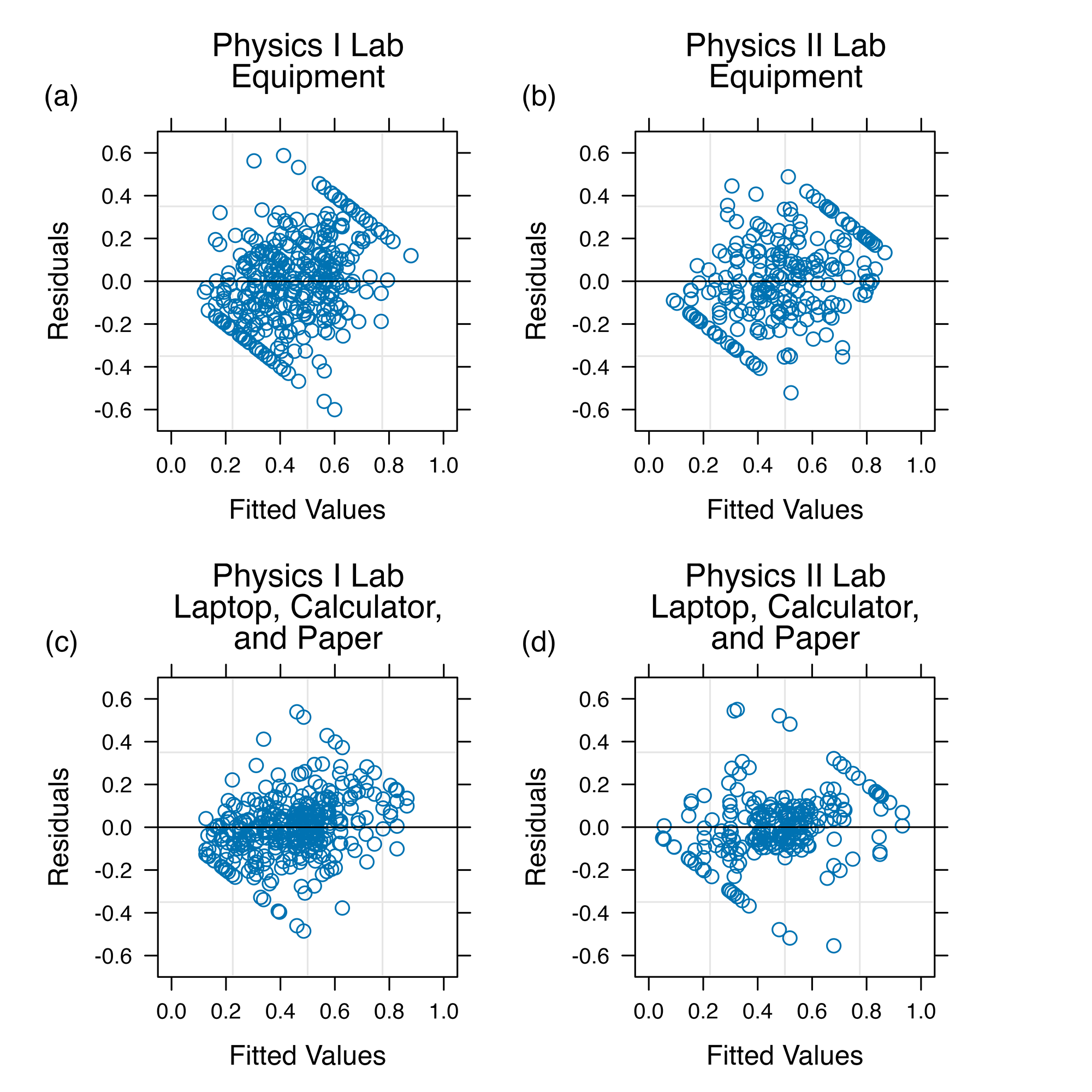}
\caption{\label{HLM_Linearity} Visual check for the assumption of linearity for hierarchical linear modeling. These plots show the residuals vs. the fitted values for a) `Equipment' group fraction in Physics I Lab, b) `Equipment' group fraction in Physics II Lab, c) `Laptop,' `Calculator,' and `Paper' group fraction in Physics I Lab, and d) `Laptop,' `Calculator,' and `Paper' group fraction in Physics II Lab.}
\end{figure}

\begin{figure}[h]
\includegraphics[width=\columnwidth]{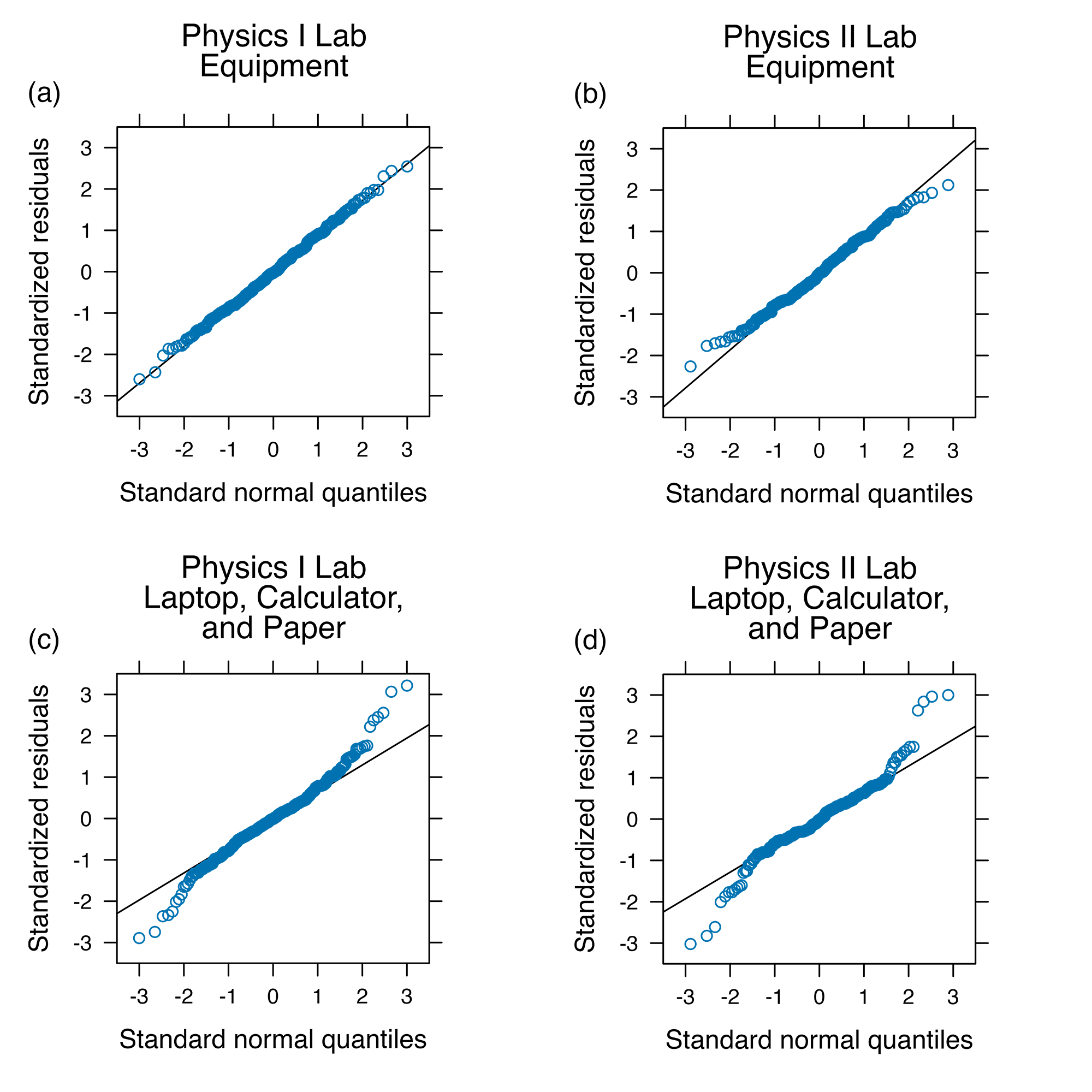}
\caption{\label{HLM_Normality} Visual check for the assumption of normality for hierarchical linear modeling. These plots show the residuals vs. the fitted values for a) `Equipment' group fraction in Physics I Lab, b) `Equipment' group fraction in Physics II Lab, c) `Laptop,' `Calculator,' and `Paper' group fraction in Physics I Lab, and d) `Laptop,' `Calculator,' and `Paper' group fraction in Physics II Lab.}
\end{figure}

In this appendix, we discuss how well assumptions were met for Hierarchical Linear Modeling (HLM). Specifically, following guidance from Van Dusen and Nissen, we examine linearity, homegeneity of variance, and normality \cite{VanDusen2019}. Examining linearity, Fig. \ref{HLM_Linearity}, we observed no trend in these data about the $x$-axes. While there are ceiling and floor effects (shown by the diagonal lines at the top right and bottom left of each plot), there is no overall structure or trend to these data as a whole. To test homogeneity of variance, ANOVA was used across all groups and students. No statistically significant differences were found ($p\sim 1$ in all four cases) indicating that variance was homogeneous. Residuals for our models are shown in Fig. \ref{HLM_Normality}. Visual inspection shows that our data generally meets the assumption of normality, falling on or near the line in all cases. While there are some deviations at either end of the plot, particularly for the third and fourth models, this does not generally impact $p$-values or estimates \cite{Lumley2002, Walsh2022}.

For our hierarchical linear modeling, we used the \verb|lmer| function from the \textit{lme4} package in R \cite{lme4Package}.

\section{Preference Survey Results}\label{sec:survey}
We present the full results of the preferences survey from Section \ref{sec:Preferences_Results} as well as the number of students for each category, Table \ref{tab:PreferencesDemographics}. Table \ref{tab:Role_Preferences} provides the roles students preferred, Table \ref{tab:RoleDistribution_Preferences} provides the role distributions students preferred, and Table \ref{tab:Leadership_Preferences} provides the leadership distributions students preferred. 

To conduct logistic regression for the roles students preferred, we used the \verb|glm| function in the \textit{base} package in R \cite{RBase}. For role distribution and leadership distribution questions, we conducted  multinomial logistic regression using the \verb|multinom| function in the \textit{nnet} package \cite{nnetPackage}. %To assess these models, we calculated McFadden pseudo-$R^2$ values using the \verb|PseudoR2| function in the \textit{DescTools} package \cite{DescToolsRPackage}.
For all of these questions, we then used the \textit{effects} package to display the probabilities of men and women selecting each answer \cite{effectsRPackage}. %For Type III ANOVA, we used the \verb|Anova| function from the \textit{car} package \cite{carPackage}. 
For pairwise comparisons of means, we used the \verb|emmeans| function from the \textit{emmeans} package \cite{emmeansPackage}.

\begin{table}[]
\caption{Number of men and women enrolled in each course and track for Physics I Lab and Physics II Lab.}
\label{tab:PreferencesDemographics}
\begin{ruledtabular}
\begin{tabular}{lcccc}
                                 & \multicolumn{2}{c}{Physics I Lab} & \multicolumn{2}{c}{Physics II Lab} \\ \cline{2-3} \cline{4-5}
Track                            & Men            & Women            & Men             & Women            \\ \hline
Algebra-based                    & 135            & 332              & 54              & 134              \\
Calculus-based     & 119            & 56               & 182             & 87               \\
\hspace{2ex}for Engineers & & & & \\
Calculus-based & 101            & 218              & 90              & 133              \\
\hspace{2ex}for Life Sciences & & & & \\
Not Enrolled      & 46             & 71               & 65              & 48              \\
\hspace{2ex}in Corequisite & & & & \\
\end{tabular}
\end{ruledtabular}
\end{table}

\onecolumngrid

\begin{sidewaystable}
% \begin{table*}[h]
\caption{Results from the logistic regressions for student role preferences given in Equation \ref{eq:PreferenceRegression} which controls for gender, course, track, and the interaction of course and track. The table shows the regression coefficient, standard error, $p$-value (in parentheses), and odds-ratio (in brackets). %The pseudo-$R^2$ values are 0.006 for the equipment model, 0.029 for the notes model, 0.014 for the analysis model, 0.026 for the managing model, and 0.013 for the no preferences model.
}
\label{tab:Role_Preferences}
\begin{ruledtabular}
\begin{tabular}{lccccc}
Predictor                                           & Equipment       & Notes           & Analysis        & Managing        & No Preference     \\ \hline
Intercept                                           & 1.033 $\pm$ 0.129  & -1.025 $\pm$ 0.125 & -0.199 $\pm$ 0.118 & 0.179 $\pm$ 0.119  & -1.946 $\pm$ 0.184   \\
                                                    & ($<0.001$)         & ($<0.001$)         & (0.091)         & (0.133)         & ($<0.001$) \\
                                                    & {[}2.809{]}     & {[}0.359{]}     & {[}0.819{]}     & {[}1.196{]}     & {[}0.143{]}       \\
Gender (ref: Man)                                   &                 &                 &                 &                 &                   \\
\hspace{2ex}Woman                                               & -0.245 $\pm$ 0.109 & 0.579 $\pm$ 0.105  & -0.322 $\pm$ 0.100 & 0.510 $\pm$ 0.100  & -0.429 $\pm$ 0.151   \\
                                                    & (0.024)         & ($<0.001$)         & (0.001)         & ($<0.001$)         & (0.004)           \\
                                                    & {[}0.782{]}     & {[}1.785{]}     & {[}0.724{]}     & {[}1.666{]}     & {[}0.651{]}       \\
Course (ref: Physics I Lab)                         &                 &                 &                 &                 &                   \\
\hspace{2ex}Physics II Lab                                      & -0.357 $\pm$ 0.182 & 0.676 $\pm$ 0.177  & 0.040 $\pm$ 0.177  & -0.349 $\pm$ 0.176 & 0.408 $\pm$ 0.264    \\
                                                    & (0.049)         & ($<0.001$)         & (0.823)         & (0.048)         & (0.121)           \\
                                                    & {[}0.700{]}     & {[}1.965{]}     & {[}1.040{]}     & {[}0.706{]}     & {[}1.504{]}       \\
Track (ref: Algebra-based)                                               &                 &                 &                 &                 &                   \\
\hspace{2ex}Calculus-based for Engineers                        & 0.022 $\pm$ 0.202  & -0.008 $\pm$ 0.197 & 0.592 $\pm$ 0.184  & -0.446 $\pm$ 0.184 & 0.076 $\pm$ 0.287    \\
                                                    & (0.915)         & (0.968)         & (0.001)         & (0.016)         & (0.791)           \\
                                                    & {[}1.022{]}     & {[}0.992{]}     & {[}1.807{]}     & {[}0.640{]}     & {[}1.079{]}       \\
\hspace{2ex}Calculus-based for Life Sciences                    & 0.026 $\pm$ 0.160  & 0.017 $\pm$ 0.153  & 0.134 $\pm$ 0.148  & -0.162 $\pm$ 0.150 & -0.006 $\pm$ 0.246   \\
                                                    & (0.873)         & (0.911)         & (0.364)         & (0.280)         & (0.980)           \\
                                                    & {[}1.026{]}     & {[}1.017{]}     & {[}1.144{]}     & {[}0.850{]}     & {[}0.994{]}       \\
\hspace{2ex}No Coreq                                            & 0.096 $\pm$ 0.231  & 0.266 $\pm$ 0.214  & 0.275 $\pm$ 0.209  & -0.262 $\pm$ 0.211 & 0.020 $\pm$ 0.344    \\
                                                    & (0.677)         & (0.214)         & (0.189)         & (0.214)         & (0.954)           \\
                                                    & {[}1.101{]}     & {[}1.305{]}     & {[}1.316{]}     & {[}0.769{]}     & {[}1.020{]}       \\
Course * Track (ref: Physics I Lab * &                 &                 &                 &                 &                   \\
\hspace{24ex}Algebra-based) &                 &                 &                 &                 &                 \\
\hspace{2ex}Physics II Lab *        & 0.317 $\pm$ 0.283  & -0.580 $\pm$ 0.276 & -0.215 $\pm$ 0.264 & 0.247 $\pm$ 0.264  & -0.158 $\pm$ 0.392   \\
\hspace{4ex}Calculus-based for Engineers	& (0.264)         & (0.035)         & (0.414)         & (0.349)         & (0.686)           \\
						& {[}1.372{]}     & {[}0.560{]}     & {[}0.806{]}     & {[}1.280{]}     & {[}0.853{]}       \\
\hspace{2ex}Physics II Lab *    & 0.209 $\pm$ 0.263  & 0.014 $\pm$ 0.252  & 0.173 $\pm$ 0.249  & -0.257 $\pm$ 0.251 & 0.069 $\pm$ 0.375    \\
\hspace{4ex}Calculus-based for Life Sciences	& (0.427)         & (0.957)         & (0.488)         & (0.305)         & (0.854)           \\
						& {[}1.232{]}     & {[}1.014{]}     & {[}1.189{]}     & {[}0.773{]}     & {[}1.071{]}       \\
\hspace{2ex}Physics II Lab *                            & 0.014 $\pm$ 0.341  & -0.435 $\pm$ 0.323 & -0.210 $\pm$ 0.320 & 0.414 $\pm$ 0.321  & -0.118 $\pm$ 0.487   \\
\hspace{4ex}No Coreq				& (0.967)         & (0.178)         & (0.510)         & (0.197)         & (0.809)           \\
						& {[}1.014{]}     & {[}0.647{]}     & {[}0.810{]}     & {[}1.513{]}     & {[}0.889{]}      
\end{tabular}
\end{ruledtabular}
% \end{table*}
\end{sidewaystable}
\twocolumngrid

\clearpage

%\begin{table*}[h]
\begin{sidewaystable}
\caption{Results from the multinomial logistic regression for student role distribution preferences similar to Equation \ref{eq:PreferenceRegression} and detailed in Section \ref{PreferencesMethods} which controls for gender, course, track, and the interaction of course and track. The table shows the regression coefficient, standard error, and $p$-value (in parentheses). %This model has a pseudo-$R^2$ of 0.016.
}
\label{tab:RoleDistribution_Preferences}
\resizebox{\textwidth}{!}{
\begin{tabular}{lcccccccccc}
\hline 
\hline
                                                      & Different Tasks & Work Together   & No Preference   & Other           & Work Together       & No Preference       & Other               & No Preference     & Other             & Other             \\
Predictor                                             & vs. Take Turns  & vs. Take Turns  & vs. Take Turns  & vs. Take Turns  & vs. Different Tasks & vs. Different Tasks & vs. Different Tasks & vs. Work Together & vs. Work Together & vs. No Preference \\ \hline
Intercept                                             & 1.475 $\pm$ 0.222  & 1.913 $\pm$ 0.216  & 0.538 $\pm$ 0.266  & -3.461 $\pm$ 1.077 & 0.438 $\pm$ 0.133      & -0.937 $\pm$ 0.204     & -4.937 $\pm$ 1.064     & -1.375 $\pm$ 0.197   & -5.374 $\pm$ 1.063   & -3.998 $\pm$ 1.074   \\
                                                      & ($<0.001$)         & ($<0.001$)         & (0.043)         & (0.001)         & (0.001)             & ($<0.001$)             & ($<0.001$)             & ($<0.001$)           & ($<0.001$)           & ($<0.001$)           \\
Gender   (ref: Man)                                   &                 &                 &                 &                 &                     &                     &                     &                   &                   &                   \\
\hspace{2ex}Woman                                                 & -0.153 $\pm$ 0.177 & -0.247 $\pm$ 0.175 & -0.713 $\pm$ 0.212 & -0.313 $\pm$ 0.524 & -0.094 $\pm$ 0.114     & -0.560 $\pm$ 0.166     & -0.160 $\pm$ 0.507     & -0.466 $\pm$ 0.164   & -0.066 $\pm$ 0.507   & 0.400 $\pm$ 0.521    \\
                                                      & (0.387)         & (0.158)         & (0.001)         & (0.551)         & (0.411)             & (0.001)             & (0.753)             & (0.004)           & (0.896)           & (0.442)           \\
Course   (ref: Physics I Lab)                         &                 &                 &                 &                 &                     &                     &                     &                   &                   &                   \\
\hspace{2ex}Physics II Lab                                        & -0.211 $\pm$ 0.290 & -0.808 $\pm$ 0.292 & -0.324 $\pm$ 0.377 & 1.162 $\pm$ 1.251  & -0.597 $\pm$ 0.198     & -0.114 $\pm$ 0.310     & 1.374 $\pm$ 1.233      & 0.483 $\pm$ 0.312    & 1.971 $\pm$ 1.233    & 1.488 $\pm$ 1.255    \\
                                                      & (0.468)         & (0.006)         & (0.390)         & (0.353)         & (0.003)             & (0.713)             & (0.265)             & (0.121)           & (0.110)           & (0.236)           \\
Track (ref: Algebra-based)                                                &                 &                 &                 &                 &                     &                     &                     &                   &                   &                   \\
\hspace{2ex}Calculus-based for Engineers                          & -0.034 $\pm$ 0.354 & -0.060 $\pm$ 0.343 & -0.020 $\pm$ 0.425 & 2.314 $\pm$ 1.178  & -0.026 $\pm$ 0.207     & 0.014 $\pm$ 0.324      & 2.349 $\pm$ 1.146      & 0.039 $\pm$ 0.312    & 2.374 $\pm$ 1.143    & 2.334 $\pm$ 1.170    \\
                                                      & (0.924)         & (0.862)         & (0.962)         & (0.050)         & (0.900)             & (0.967)             & (0.040)             & (0.899)           & (0.038)           & (0.046)           \\
\hspace{2ex}Calculus-based for Life Sciences                      & -0.408 $\pm$ 0.252 & -0.663 $\pm$ 0.246 & -0.305 $\pm$ 0.320 & 1.305 $\pm$ 1.139  & -0.256 $\pm$ 0.166     & 0.102 $\pm$ 0.263      & 1.714 $\pm$ 1.125      & 0.358 $\pm$ 0.257    & 1.969 $\pm$ 1.124    & 1.610 $\pm$ 1.142    \\
                                                      & (0.106)         & (0.007)         & (0.339)         & (0.252)         & (0.125)             & (0.696)             & (0.128)             & (0.163)           & (0.080)           & (0.158)           \\
\hspace{2ex}No Corequisite                                              & -0.894 $\pm$ 0.330 & -1.016 $\pm$ 0.317 & -0.336 $\pm$ 0.396 & 0.613 $\pm$ 1.441  & -0.122 $\pm$ 0.251     & 0.558 $\pm$ 0.346      & 1.507 $\pm$ 1.428      & 0.681 $\pm$ 0.333    & 1.630 $\pm$ 1.425    & 0.949 $\pm$ 1.444    \\
                                                      & (0.007)         & (0.001)         & (0.397)         & (0.670)         & (0.626)             & (0.106)             & (0.291)             & (0.041)           & (0.253)           & (0.511)           \\
Course *   Track (ref: Physics I Lab * Algebra-based) &                 &                 &                 &                 &                     &                     &                     &                   &                   &                   \\
\hspace{2ex}Physics II Lab * Calculus-based for Engineers         & 0.118 $\pm$ 0.468  & 0.187 $\pm$ 0.465  & 0.584 $\pm$ 0.569  & -2.546 $\pm$ 1.556 & 0.069 $\pm$ 0.301      & 0.465 $\pm$ 0.444      & -2.666 $\pm$ 1.516     & 0.397 $\pm$ 0.441    & -2.734 $\pm$ 1.515   & -3.130 $\pm$ 1.549   \\
                                                      & (0.800)         & (0.687)         & (0.305)         & (0.102)         & (0.819)             & (0.295)             & (0.079)             & (0.368)           & (0.071)           & (0.043)           \\
\hspace{2ex}Physics II Lab * Calculus-based for Life Sciences     & 0.818 $\pm$ 0.424  & 1.222 $\pm$ 0.426  & 0.771 $\pm$ 0.537  & -1.757 $\pm$ 1.701 & 0.404 $\pm$ 0.281      & -0.047 $\pm$ 0.431     & -2.578 $\pm$ 1.671     & -0.451 $\pm$ 0.432   & -2.981 $\pm$ 1.671   & -2.531 $\pm$ 1.703   \\
                                                      & (0.054)         & (0.004)         & (0.151)         & (0.301)         & (0.150)             & (0.913)             & (0.123)             & (0.297)           & (0.074)           & (0.137)           \\
\hspace{2ex}Physics II Lab * No Corequisite                             & 1.335 $\pm$ 0.558  & 1.794 $\pm$ 0.551  & 0.967 $\pm$ 0.670  & 0.844 $\pm$ 1.755  & 0.459 $\pm$ 0.371      & -0.368 $\pm$ 0.531     & -0.491 $\pm$ 1.707     & -0.827 $\pm$ 0.523   & -0.950 $\pm$ 1.705   & -0.124 $\pm$ 1.746   \\
                                                      & (0.017)         & (0.001)         & (0.149)         & (0.631)         & (0.216)             & (0.488)             & (0.774)             & (0.114)           & (0.577)           & (0.943)           \\ \hline \hline         
\end{tabular}
}
\end{sidewaystable}
%\end{table*}

%\clearpage
\begin{sidewaystable}
%\begin{table*}[h]
\caption{Results from the multinomial logistic regression for student leadership preferences similar to Equation \ref{eq:PreferenceRegression} and detailed in Section \ref{PreferencesMethods} which controls for gender, course, track, and the interaction of course and track. The table shows the regression coefficient, standard error, and $p$-value (in parentheses).
\label{tab:Leadership_Preferences}} %This model has a pseudo-$R^2$ of 0.016.
\resizebox{\textwidth}{!}{
\begin{tabular}{lcccccccccc}
\hline 
\hline
                                                      & Take Turns      & One Leader      & No Preference   & Other            & One Leader      & No Preference   & Other            & No Preference   & Other            & Other             \\
Predictor                                             & vs. No Leader   & vs. No Leader   & vs. No Leader   & vs. No Leader    & vs. Take Turns  & vs. Take Turns  & vs. Take Turns   & vs. One Leader  & vs. One Leader   & vs. No Preference \\ \hline
Intercept                                             & 0.962 $\pm$ 0.178  & 0.432 $\pm$ 0.195  & 0.098 $\pm$ 0.212  & -1.677 $\pm$ 0.405  & -0.530 $\pm$ 0.149 & -0.864 $\pm$ 0.171 & -2.640 $\pm$ 0.385  & -0.334 $\pm$ 0.189 & -2.110 $\pm$ 0.394  & -1.776 $\pm$ 0.402   \\
                                                      & ($<0.001$)         & (0.027)         & (0.643)         & ($<0.001$)          & ($<0.001$)         & ($<0.001$)         & ($<0.001$)          & (0.077)         & ($<0.001$)          & ($<0.001$)           \\
Gender   (ref: Man)                                   &                 &                 &                 &                  &                 &                 &                  &                 &                  &                   \\
\hspace{2ex}Woman                                                 & 0.467 $\pm$ 0.153  & 0.205 $\pm$ 0.165  & -0.102 $\pm$ 0.175 & 0.194 $\pm$ 0.384   & -0.262 $\pm$ 0.127 & -0.569 $\pm$ 0.141 & -0.272 $\pm$ 0.369  & -0.307 $\pm$ 0.154 & -0.010 $\pm$ 0.375  & 0.297 $\pm$ 0.379    \\
                                                      & (0.002)         & (0.215)         & (0.559)         & (0.613)          & (0.040)         & ($<0.001$)         & (0.460)          & (0.046)         & (0.978)          & (0.434)           \\
Course   (ref: Physics I Lab)                         &                 &                 &                 &                  &                 &                 &                  &                 &                  &                   \\
\hspace{2ex}Physics II Lab                                        & -0.259 $\pm$ 0.264 & -0.242 $\pm$ 0.295 & 0.192 $\pm$ 0.311  & 0.157 $\pm$ 0.522   & 0.017 $\pm$ 0.228  & 0.451 $\pm$ 0.249  & 0.416 $\pm$ 0.487   & 0.434 $\pm$ 0.282  & 0.399 $\pm$ 0.505   & -0.035 $\pm$ 0.514   \\
                                                      & (0.326)         & (0.412)         & (0.537)         & (0.763)          & (0.941)         & (0.071)         & (0.392)          & (0.124)         & (0.429)          & (0.946)           \\
Track (ref: Algebra-based)                                                &                 &                 &                 &                  &                 &                 &                  &                 &                  &                   \\
\hspace{2ex}Calculus-based for Engineers                          & 0.027 $\pm$ 0.275  & -0.223 $\pm$ 0.313 & 0.037 $\pm$ 0.331  & -0.027 $\pm$ 0.595  & -0.250 $\pm$ 0.241 & 0.011 $\pm$ 0.266  & -0.053 $\pm$ 0.561  & 0.260 $\pm$ 0.305  & 0.196 $\pm$ 0.581   & -0.064 $\pm$ 0.591   \\
                                                      & (0.923)         & (0.476)         & (0.911)         & (0.964)          & (0.301)         & (0.968)         & (0.924)          & (0.393)         & (0.736)          & (0.913)           \\
\hspace{2ex}Calculus-based for Life Sciences                      & 0.344 $\pm$ 0.245  & 0.358 $\pm$ 0.266  & 0.422 $\pm$ 0.292  & -0.496 $\pm$ 0.613  & 0.014 $\pm$ 0.182  & 0.078 $\pm$ 0.218  & -0.840 $\pm$ 0.581  & 0.065 $\pm$ 0.242  & -0.854 $\pm$ 0.591  & -0.918 $\pm$ 0.603   \\
                                                      & (0.160)         & (0.179)         & (0.147)         & (0.418)          & (0.940)         & (0.720)         & (0.149)          & (0.790)         & (0.149)          & (0.128)           \\
\hspace{2ex}No Corequisite                                              & 0.198 $\pm$ 0.342  & 0.181 $\pm$ 0.374  & 0.482 $\pm$ 0.394  & -11.810 $\pm$ 0.433 & -0.017 $\pm$ 0.263 & 0.284 $\pm$ 0.292  & -12.450 $\pm$ 0.416 & 0.301 $\pm$ 0.329  & -11.823 $\pm$ 0.423 & -12.099 $\pm$ 0.424  \\
                                                      & (0.562)         & (0.628)         & (0.220)         & ($<0.001$)          & (0.949)         & (0.330)         & ($<0.001$)          & (0.360)         & ($<0.001$)          & ($<0.001$)           \\
Course *   Track (ref: Physics I Lab * Algebra-based) &                 &                 &                 &                  &                 &                 &                  &                 &                  &                   \\
\hspace{2ex}Physics II Lab * Calculus-based for Engineers         & -0.256 $\pm$ 0.391 & 0.334 $\pm$ 0.436  & -0.110 $\pm$ 0.456 & -0.996 $\pm$ 0.884  & 0.590 $\pm$ 0.346  & 0.146 $\pm$ 0.371  & -0.741 $\pm$ 0.843  & -0.444 $\pm$ 0.419 & -1.330 $\pm$ 0.865  & -0.886 $\pm$ 0.875   \\
                                                      & (0.512)         & (0.445)         & (0.809)         & (0.260)          & (0.088)         & (0.695)         & (0.380)          & (0.289)         & (0.124)          & (0.312)           \\
\hspace{2ex}Physics II Lab * Calculus-based for Life Sciences     & -0.636 $\pm$ 0.385 & -0.113 $\pm$ 0.417 & -0.557 $\pm$ 0.447 & -1.732 $\pm$ 1.257  & 0.523 $\pm$ 0.316  & 0.079 $\pm$ 0.356  & -1.097 $\pm$ 1.228  & -0.444 $\pm$ 0.390 & -1.619 $\pm$ 1.238  & -1.174 $\pm$ 1.249   \\
                                                      & (0.098)         & (0.786)         & (0.212)         & (0.168)          & (0.098)         & (0.825)         & (0.372)          & (0.255)         & (0.191)          & (0.347)           \\
\hspace{2ex}Physics II Lab * No Corequisite                             & -0.301 $\pm$ 0.501 & 0.008 $\pm$ 0.546  & -0.201 $\pm$ 0.560 & 11.111 $\pm$ 0.433  & 0.309 $\pm$ 0.414  & 0.100 $\pm$ 0.434  & 11.854 $\pm$ 0.416  & -0.209 $\pm$ 0.484 & 10.935 $\pm$ 0.423  & 11.119 $\pm$ 0.424   \\
                                                      & (0.549)         & (0.988)         & (0.720)         & ($<0.001$)          & (0.456)         & (0.818)         & ($<0.001$)          & (0.666)         & ($<0.001$)          & ($<0.001$)    \\ \hline \hline 
\end{tabular}
}
%\end{table*}
\end{sidewaystable}

%\section{Video Observations Regression Results}\label{sec:VideoAnalysisRegression}
%In this section, we share the full results of our regression models for video analysis. For this hierarchical linear modeling, we used the \verb|lmer| function from the \textit{lme4} package in R \cite{lme4Package}. 

\clearpage
\bibliography{refs}
\end{document}